\documentclass{optica-article}

\journal{opticajournal} 

\articletype{Research Article}

\usepackage{array}
\usepackage{fancyhdr}
\usepackage{subcaption}
\usepackage{siunitx}
\usepackage{rotating}
\usepackage[table]{xcolor}
\usepackage{enumitem}
\newlist{balditemize}{itemize}{1}
\setlist[balditemize]{label=, wide=\parindent, labelsep*=0pt, leftmargin=*, topsep=0pt, itemsep=0pt}
\usepackage{caption}
\usepackage{hyperref}

\usepackage{lineno}

\begin{document}

\title{Tri-coupler geometries for achromatic nulling interferometry in the near-infrared}

\author{Harry-Dean Kenchington Goldsmith,\authormark{1,*} Nemanja Jovanovic,\authormark{2} Anusha Pai Asnodkar,\authormark{2} Yoo Jung Kim,\authormark{3} Ahmed Sanny,\authormark{3} Pradip Gatkine\authormark{3} and Michael P. Fitzgerald\authormark{3}}

\address{
\authormark{1}Astrophotonics Australia, Canberra, ACT, Australia\\
\authormark{2}{Department of Astronomy}, {California Institute of Technology}, {{1200 East California Boulevard}, {Pasadena}, {91125}, {CA}, {USA}}\\
\authormark{3}Physics \& Astronomy Department, University of California, Los Angeles, 475 Portola Plaza, Los Angeles, 90095, CA, USA\\
}

\email{\authormark{*}harrykenchington@astrophotonics.com.au} 


\begin{abstract*} 
Astrophotonics will be central to the next generation of astronomical instrumentation, enabling lightweight, compact, and environmentally stable photonic integrated circuits for both ground-based observatories and future space missions. One key application is beam combination for nulling interferometry, which suppresses starlight to reveal exoplanets and companions. Compact, broadband photonic beam combiners are essential for enabling complex circuitry on a single chip and for scalable solutions for single- and multi-telescope instruments, and are investigated herein.

Two-waveguide photonic combiners rely on symmetric evanescent coupling to interfere light, which is inherently chromatic and requires modification for broadband operation. A three-waveguide configuration, or tri-coupler, offers the potential for deeper, broader, and more stable achromatic nulls compared with two-waveguide approaches. This work compares the simulated performance of two evanescent tri-couplers and a multimode interference coupler~(MMI) across the 1.5--1.8\,\si{\um} band, evaluating exoplanet throughput, starlight attenuation, sensing characteristics, and estimations on fabrication tolerance. All three tri-couplers achieved $>40$\,dB attenuation over a $\geq270$\,nm bandwidth. However, the standard tri-coupler was outperformed by both a bespoke tapered tri-coupler and the MMI, each of which achieved exoplanet throughput $>85$\% across the band, excluding component losses. Including component loss, the tapered tri-coupler has the highest total throughput, averaging $\sim$96\%. The standard tri-coupler began with an equivalent exoplanet throughput, falling to 50\% at the band edges. The tapered tri-coupler was further redesigned to achieve a non-degenerate sensing state. The MMI, while limited to a starlight attenuation of 40\,dB $\left(10^{-4}\right)$ by uncoupled light, showed the greatest tolerance to fabrication errors, offering strong practical potential.

Future designs aim to combine high exoplanet throughput, deep starlight attenuation, and non-degenerate sensing within a single integrated architecture. This work provides a simulation suite for three tri-couplers. They can be selected based on robustness to common fabrication tolerances (the MMI), exoplanet throughput (the tapered tri-coupler), and/or the sensing performance (the tapered tri-coupler).
\end{abstract*}

\section{Introduction}

The first exoplanet was detected in 1992~\cite{Wolszczan1992} using an indirect detection method --- pulsar timing. Since then, $>6100$ exoplanets have been confirmed (as of March 2026 from NASA's exoplanet archive~\cite{nasa_exoplanet_archive2025}) predominantly via indirect techniques such as the transit and radial velocity methods. Direct detection approaches, including coronagraphy and interferometry, account for only a fraction of detections and typically identify large, young exoplanets and brown dwarfs that remain luminous due to their thermal emission~\cite{defrere2022}. The closer an exoplanet orbits its host star, the more difficult it becomes to disentangle planetary light from the overwhelming stellar glare. As a result, direct imaging is currently most effective for exoplanets at wide separations — tens to thousands of astronomical units. To probe closer orbits and detect fainter companions, new starlight suppression techniques are required.

Starlight suppression has two main pathways: coronagraphy and nulling interferometry. Coronagraphy requires blocking the on-axis signal~\cite{Lyot1939} to observe fainter off-axis sources. Obtaining a small inner working angle is challenging, which limits observations of close companion exoplanets~\cite{Guyon2006}.

In contrast, nulling interferometry employs destructive interference between two beams from segmented or multiple telescopes. Beam combination creates bright and dark fringes corresponding to phase differences across the sky. A 180° phase shift between telescopes suppresses on-axis starlight while maintaining off-axis sources, such as exoplanets~\cite{Bracewell1978}. 

Nulling applications include exoplanet detection and characterization. The required starlight suppression depends on the exoplanet investigated. For example, the planet-star flux ratio for a close-orbit hot Jupiter or young, self-luminous, gas giant still in the process of forming is $\sim10^{-4}$~\cite{Marois2006}, whereas a cool Earth-like potentially habitable exoplanet is $10^{-10}$ at a wavelength of 1.55\,\si{\um}.
High contrast instrumentation is also needed for direct detection of close-orbit low mass stars and brown dwarfs ($10^{-2}$)~\cite{Echeverri2024}. For astronomical nulling instrumentation, Photonic Integrated Circuits~(PICs) offer a unique opportunity. They use photonic beam combiners on a miniaturized, monolithic chip to null starlight and reveal objects around the star.

Astrophotonics~\cite{Bland-Hawthorn2009, Bryant2017, Dinkelaker2021, Jovanovic2023} is an emerging field that leverages photonic technologies developed by universities, research centers, and the telecommunications industry to manipulate light from a telescope in waveguides and fibers before a science camera. These devices enable on-chip starlight suppression for direct exoplanet detection. Photonic beam combiners are a natural candidate for nulling interferometry because they can separate starlight and surrounding light, such as exoplanet light~\cite{Errmann2015}. There are many examples of PICs at observatories~\cite{Gillessen2010,Perrin2006,Huby2012,Norris2014, Setterholm2023}, but only a handful are used for nulling interferometry. The Guided Light Interferometric Nulling Technologies~(GLINT)~\cite{Norris2014}  at the Subaru telescope uses beam combiners for nulling interferometry in the near-infrared~\cite{Arcadi2024} and the Nulling Observations of exoplaneTs and dusT~(NOTT) instrument of the Asgard instrument suit is a mid-infrared nulling interferometer~\cite{martinod2023}  installed at the Very Large Telescope Interferometer~(VLTI) visitor instrument bench~\cite{Sanny2026}. When fabricating these photonic chips, the choice of beam combiner is a critical step in PIC design, with coupler selection restricted by materials and fabrication methods.  

Two-input-two-output beam combiners, based on asymmetric directional couplers~\cite{Gretzinger2019} and Multimode Interference couplers (MMI)~\cite{KenchingtonGoldsmith2016, KenchingtonGoldsmith2018, Morley2016}, have been demonstrated as broadband components for nulling interferometry.  MMIs are useful for their tolerance to fabrication errors~\cite{Besse1994}, making them competitive with evanescent couplers, though typically a more achromatic nulling performance comes at the cost of lower throughput~\cite{KenchingtonGoldsmith2017b}. Tapered evanescent couplers have been suggested~\cite{KenchingtonGoldsmith2024} too, but the gold standard seems to be the equidistant evanescent tri-coupler~\cite{Martinod:21}, as is used on GLINT~\cite{Arcadi2024}. Both NOTT and GLINT utilized ultrafast laser inscription~(ULI) in bulk glass: a low-index-contrast technology ($\Delta\text{n}\approx0.001$) that enables low-loss, three-dimensional waveguides~\cite{Gretzinger2015}. However, like with all technology, there are trade-offs. This technology is restricted by its large bend radii, limiting scalability. If more components are required to service more telescopes, for example, it may not be suitable.

High-index-contrast material systems ($\Delta\text{n}>0.1$) fabricated via thin-film lithography enable compact two-dimensional waveguides with tight bend radii, enhancing scalability. However, these often suffer from higher coupling losses due to mode mismatch with standard optical fibers. Upcoming facilities such as the Giant Magellan Telescope~(GMT)~\cite{Bernstein2014} and the Habitable Worlds Observatory~(HWO)~\cite{Dressing2024Habitable} will require compact, densely integrated photonic circuits, making high-index-contrast technologies increasingly attractive.

Here, we investigate work using lithographically fabricated waveguides --- a silica substrate with a deposited germanium-doped silica core layer and a silica top layer. At an index ratio of 2\%, the waveguides have efficient fiber coupling of approximately 95\%, minimized dispersion, low material loss, and polarization independence when employing square waveguides~\cite{Gatkine2024}. With this platform, both evanescent and MMI-based couplers are viable candidates.

Following previous work on nulling interferometry for three-dimensional tri-couplers~\cite{Arcadi2024}, two-dimensional tri-couplers are evaluated in this work for application in nulling interferometry. While typically consisting of three evanescently coupled waveguides, the term tri-coupler in this work is extended to include any three-input-three-output devices, such as a 3$\times$3~MMI. Three designs are compared in this work (Sec.\,\ref{sec:tricouplers}): a standard tri-coupler with three equal-width waveguides coupled evanescently, a custom-made 3$\times$3~MMI, and a bespoke tapered tri-coupler inspired by Hsiao~et~al.~\cite{Hsiao2009}. 

The comparative performance of these devices is evaluated in terms of exoplanet throughput  (Sec.\,\ref{sec:Throughput}), starlight suppression (Sec.\,\ref{sec:Null}, component losses (Sec.\,\ref{sec:Loss}) including bend loss, and sensing (Sec.\,\ref{sec:Sensing}). Null depth limitations, fabrication tolerances, and the resulting starlight suppression are discussed in Sec.\,\ref{sec:Null}. These metrics provide a realistic assessment of these tri-couplers as nulling beam combiners and their expected on-chip performance.

\section{Tri-couplers}
\label{sec:tricouplers}

\begin{figure}[ht]
    \centering
        \begin{subfigure}{0.32\textwidth}
        \centering
               \includegraphics[width=2.0\linewidth, angle=90]{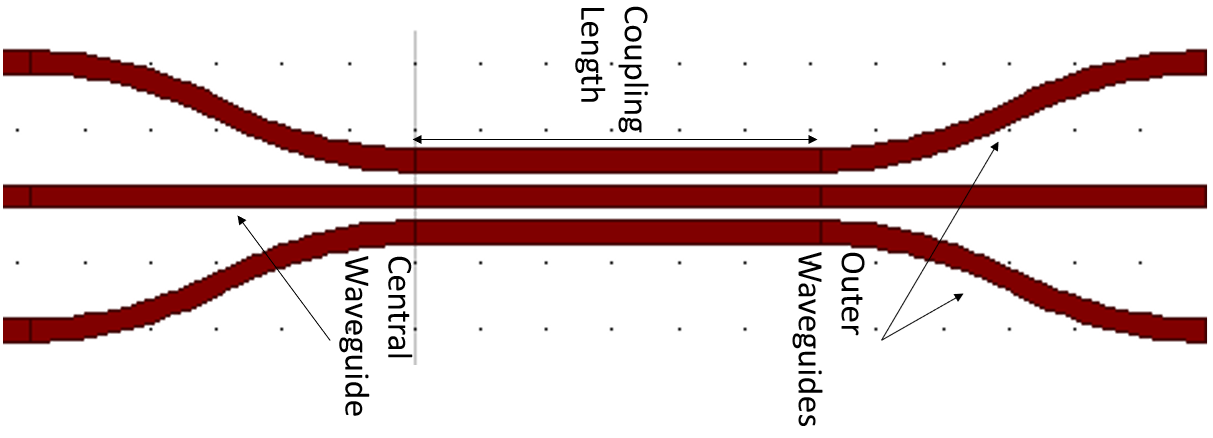}
            \caption{Standard tri-coupler}
            \label{fig:CAD_Tricoupler}
        \end{subfigure}
        \begin{subfigure}{0.32\textwidth}    \centering
                \hspace{20pt}\includegraphics[ width=2.0\linewidth, angle=90]{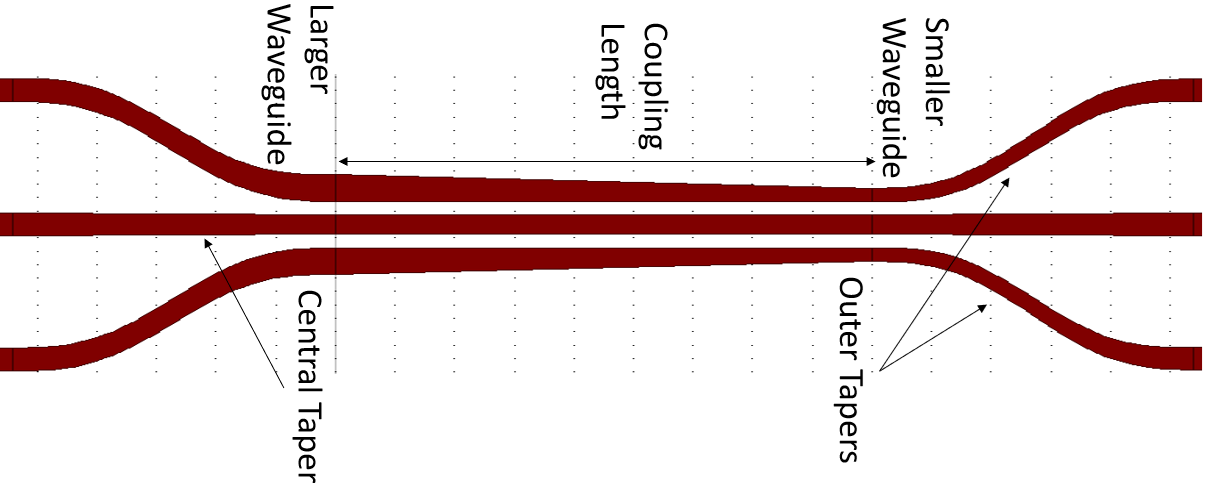}
            \caption{Tapered tri-coupler}
            \label{fig:CAD_TaperedTricoupler}
        \end{subfigure}
        \begin{subfigure}{0.32\textwidth}     \centering
        
                \hspace{10pt}\includegraphics[width=2.0\linewidth, angle=90]{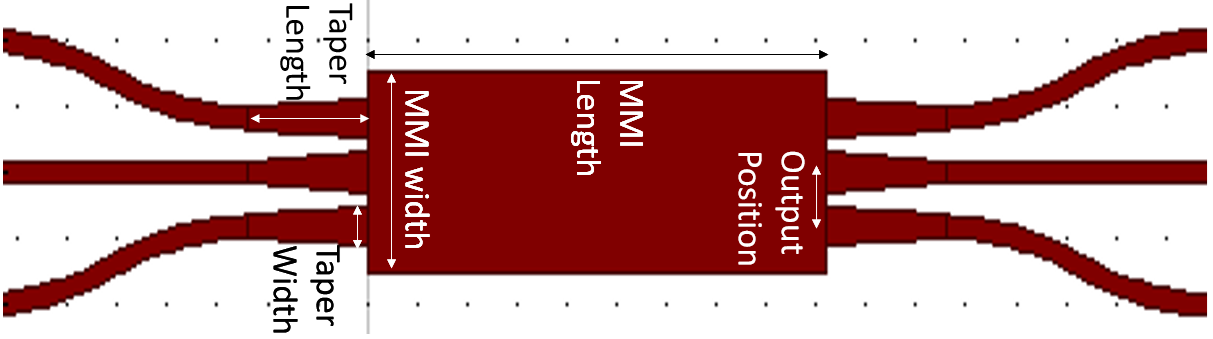}
            \caption{MMI}
            \label{fig:CAD_MMI}
        \end{subfigure}
    \caption{The RSoft CAD of the three tri-couplers (not to scale).}
    \label{fig:CAD}
\end{figure}

Three tri-coupler configurations were investigated for use in on-chip nulling interferometry: a standard evanescent tri-coupler, a tapered tri-coupler, and an MMI. The three Computer-Aided Designs~(CADs) used for this work are shown in Fig.\,\ref{fig:CAD}.

RSoft CAD and BeamPROP (version 2021.09) are interactive photonic design and beam propagation simulation programs, respectively~\cite{RSOFT}, and were used to simulate the tri-couplers in this work. The default waveguide size, used at the input and output of each device, was 3.4$\times$3.4\,\si{\um}. These waveguides were simulated as completely square. Lithographically fabricated waveguide sidewalls have an expected $\sim$1° sidewall deviation, with minor impact on performance and birefringence. Previous studies have concluded that the birefringence of these waveguides is negligible~\cite{Gatkine2024}. 

Each simulation was undertaken using transverse electric (TE) mode fields in a 3-D semi-vector configuration. Full-vector simulations were not used because polarization transfer was not expected. We used the built-in simulation material (SiO2: n-index = 1.444, k-index = 0 at 1.55\,\si{\um}) for the cladding material, which was lossless over the wavelength range, with a 2\% index contrast to the fully etched, fully immersed core material. At these wavelengths, the default waveguide is single-mode with an effective index of 1.458 at 1.55\,\si{\um}, but, as will be shown, the simulated components may include multimode components. This model and these parameters were used to be consistent with previous work~\cite{Gatkine2024}. The bend radius --- Cosine S-bends for lower losses at a small length~\cite{Kruse2015} --- was set to 850\,\si{\um} to minimize the tri-couplers' footprint, with a minimal loss at 1.55\,\si{\um} at 3.5\% loss (0.16\,dB) per bend junction, and, as will be discussed in Sec.\,\ref{sec:Loss}, is too high a loss for a nulling interferometer. 

All three tri-couplers have three input waveguides and three output waveguides (one central and two outer). In practice, only the outer input waveguides are routed to the edge of the chip for star/planet light injection (the center input is only used for testing purposes), and all three output waveguides are routed to the chip's end for detection. The existence of all three input waveguides is critical in the interaction regions of the standard and tapered tri-coupler variants due to evanescent coupling occurring in the bends.  

\subsection{Nulling interferometry using tri-couplers}
The tri-coupler suppresses starlight at the device's central output. This occurs in a two-dimensional tri-coupler only when equal-intensity electric fields with a 180° phase difference are injected into the outer input ports. This choice is agnostic for three-dimensional symmetrical tri-couplers. These simulations assume achromatic phase shifters~\cite{Douglass2025} have been placed before the tri-couplers. 

All light will undergo this phase difference whilst propagating to the tri-coupler, but objects spatially separated from the star will have a phase offset. This is because an exoplanet is off-axis in the focal plane; its beam has a tilted wavefront in the pupil plane and thus a geometric path difference from the star between the telescopes. An exoplanet at an angular separation of $\lambda/2D$, for example, where $\lambda$ is the observing wavelength and $D$ denotes either the telescope diameter or the interferometric baseline length, introduces an extra 180° (i.e., an optical delay of $\lambda/2$) phase difference compared to the on-axis starlight (without a phase tilt). For this work, we simulate a $\lambda/2D$ separation using a simplified two-telescope model: two equal TE fields input into the tri-couplers' outer waveguides, and an achromatic 180° phase difference between exoplanet and star light, producing constructive and destructive interference into the tri-coupler's central waveguide, respectively.

The simulations conducted in this work identified the design parameters for a starlight attenuation, defined as:

\begin{equation}
    \text{Attenuation (dB)} =\eta_S= -10~\log_{10}\left(\frac{\text{Destructive output}}{ \sum\text{All outputs}}\right)
    \label{eq:Attenutation}
\end{equation}
which excludes exoplanet light throughput ($\eta_P$) and is distinct from the null depth (the ratio between destructive and constructive outputs). To understand how these metrics determine exoplanet detection, we consider the signal-to-noise ($S/N$) ratio of a starlight suppression device. 

\begin{equation}
    \text{Signal-to-noise ratio (SNR)} \propto \frac{\text{S}_{\text{p}}}{\sqrt{\text{S}_{\text{s}}}} \propto\frac{\eta_{\text{p}}}{\sqrt{\eta_{\text{s}}}} 
    \label{eq:sn}
\end{equation}

that directly compares the signal from the exoplanet ($\text{S}_{\text{p}}$) and the star ($\text{S}_{\text{s}}$) and the PIC throughput~($\eta$)~\cite{Ruane2018}. Hence, from Eq.\,\eqref{eq:sn}, the SNR is more sensitive to changes in the exoplanet throughput than starlight suppression. Therefore, exoplanet throughput should be the primary driving requirement when optimizing tri-couplers. Starlight attenuation is also important and should be maximized without compromising exoplanet throughput. The null depth typically used in nulling interferometry~\cite{Serabyn1999}, and other high contrast imaging, is defined as
\begin{equation}
    \text{Null depth (dB)} = -10~\log_{10}\left(\frac{\text{Destructive output}}{ \text{Constructive output}}\right)
    \label{eq:null}
\end{equation}
To be clear, this equation compares the light in the tri-coupler's central output port when injecting light with a 0° phase difference (the exoplanet's constructive interference output) and 180° phase difference (the star light's destructive interference output).

After the PIC, the star-to-planet contrast can be increased through self-calibration techniques~\cite{Hanot2011,Sanny2026} or using high-dispersion spectroscopy~\cite{Snellen2015} up to the photon noise limit. The implication is that having a complete attenuation of the star in the PIC is not necessary when post-processing is available. Thus, the focus of this paper is a comparison between three tri-couplers' abilities to transmit exoplanet light, attenuate starlight, and overall loss rather than stipulating requirements for observations. 

\subsection{Tri-coupler designs for nulling interferometry}
\label{sec:TricouplerDesigns}

This work prioritizes exoplanet throughput (including system loss) followed by starlight suppression. However, to stabilize the null, fringe tracking is required, making fringe tracking and starlight suppression equally important. The optimization process for each tri-coupler focused on directing exoplanet light to the central output port at 1.55\,\si{\um}, with fringe tracking secondary. 

\begin{figure}[htbp]
    \centering
    \setlength{\tabcolsep}{1pt}
    \renewcommand{\arraystretch}{1.3}

    \begin{tabular}{@{} c c c c @{}}
        & \hspace{64pt} \textbf{Split} & \hspace{-20pt} \textbf{Constructive} &  \hspace{-100pt}\textbf{Destructive} \\[4pt]

        \multirow{3}{*}[4.0cm]{\centering\rotatebox{90}{\textbf{Standard tri-coupler}}} &
        \begin{subfigure}{0.3\textwidth}
            \centering
            \raisebox{-0.9cm}{\hspace*{0.95cm}\includegraphics[width=3.45cm,height=5.3cm]{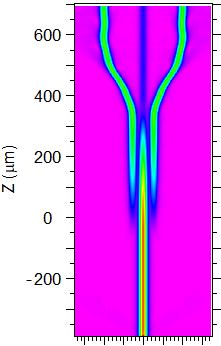}}
            \label{fig:Tricoupler_Split}
        \end{subfigure} &
        \begin{subfigure}{0.3\textwidth}
            \centering
             \raisebox{0.03cm}
            {\hspace*{-0.55cm}\includegraphics[width=2.45cm,height=5.3cm]{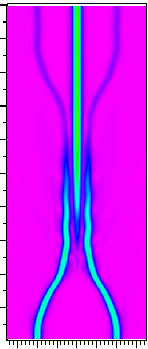}}
            \label{fig:Tricoupler_AntiNull}
        \end{subfigure} &
        \begin{subfigure}{0.3\textwidth}
            \centering
            \hspace*{-3.6cm}\includegraphics[width=2.40cm,height=5.25cm]{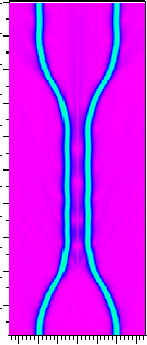}
            \label{fig:Tricoupler_Null}
        \end{subfigure} \\[-4pt]
       \multirow{3}{*}[3.2cm]{\centering\rotatebox{90}{\textbf{MMI}}} &
        \begin{subfigure}{0.3\textwidth}
            \centering
            \raisebox{-0.5cm}{\hspace*{0.7cm}\includegraphics[width=3.75cm,height=5.3cm]{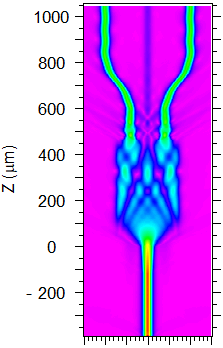}}
            \label{fig:MMI_Split}
        \end{subfigure} &
        \begin{subfigure}{0.3\textwidth}
            \centering
            \raisebox{0.05cm}
             {\hspace*{-0.55cm}\includegraphics[width=2.55cm,height=5.3cm]{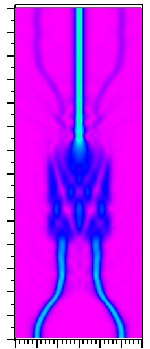}}
            \label{fig:MMI_AntiNull}
        \end{subfigure} &
        \begin{subfigure}{0.3\textwidth}
            \centering
            \raisebox{0.05cm}
            {\hspace*{-3.6cm}\includegraphics[width=2.5cm,height=5.23cm]{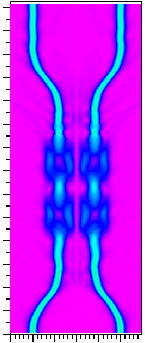}}
            \label{fig:MMI_Null}
        \end{subfigure} \\[-8pt]
        \multirow{3}{*}[4.6cm]{\centering\rotatebox{90}{\textbf{Tapered tri-coupler}}} &
        \begin{subfigure}{0.3\textwidth}
            \centering
            \raisebox{-0.5cm}{\hspace*{1.0cm}\includegraphics[width=3.5cm,height=6.05cm]{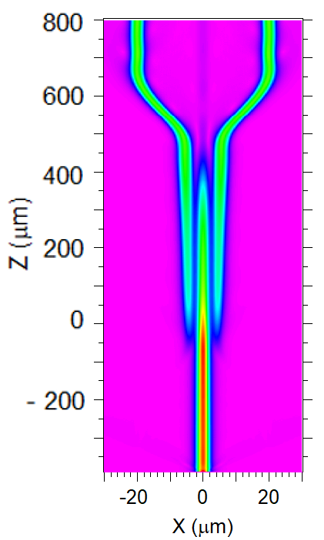}}
            \label{fig:TaperedTricoupler_Split}
        \end{subfigure} &
        \begin{subfigure}{0.3\textwidth}
            \centering
            {\hspace*{-0.5cm}\includegraphics[width=2.4cm,height=5.9cm]{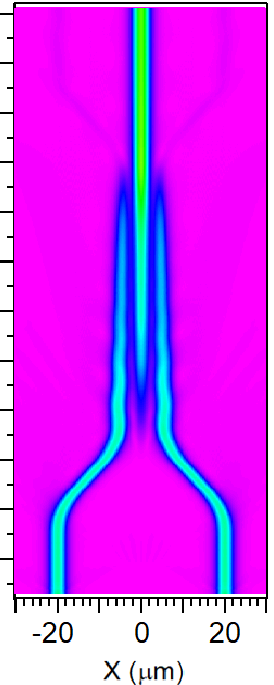}}
            \label{fig:TaperedTricoupler_AntiNull}
        \end{subfigure} &
        \begin{subfigure}{0.3\textwidth}
            \centering
            \hspace*{-3.6cm}\includegraphics[width=2.4cm,height=5.9cm]{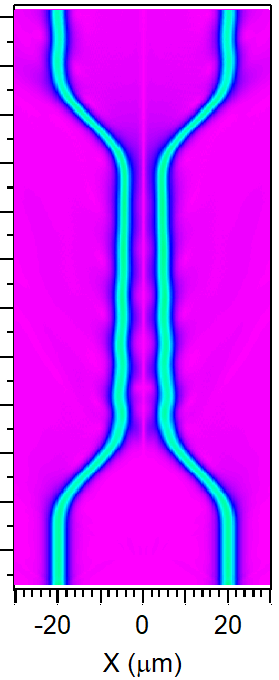}
            \label{fig:TaperedTricoupler_Null}
        \end{subfigure}\\
        &&\includegraphics[width=0.3\linewidth]{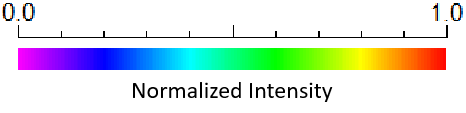}
    \end{tabular}
    \caption{Normalized intensity in the three tri-coupler configurations (Standard, MMI, and Tapered) for the splitting of a single input (Split), the constructive interference in the coupler for two equal intensity inputs with a 0° phase offset (Constructive - exoplanet light path), and the destructive interference for two equal intensity inputs with a 180° phase offset (Destructive - star light path).}
    \label{fig:BP_Examples}
\end{figure}

A simple method for identifying the exoplanet's throughput is to run the couplers in reverse: injecting light into the central input port and simulating the 50:50 splitting ratio. The length at which equal splitting is obtained was the length selected for this work. Future experimental verification will use this as a quick method to test exoplanet throughput and achromaticity. Examples of the splitting ratio, the maximized exoplanet light into the central output port, and the starlight attenuation in the central output for each tri-coupler are shown in Fig.\,\ref{fig:BP_Examples}.

\subsubsection{Standard tri-coupler}

The standard tri-coupler length is equivalent to calculating the beat length~($L_{\pi}$), which is the length at which the initial field distribution completely transfers to the other waveguides --- light in the central waveguide splitting into the two outer waveguides, for example. 
For light injecting into the central waveguide, $L_{\pi}$ is determined by the difference in effective indices of the symmetric modes, and given by~\cite{Donnelly1986}:
\begin{equation}
    L_{\pi}=\frac{\lambda}{n_A-n_B}
    \label{eq:BeatLength}
\end{equation}
where $n_A$ and $n_B$ are the effective indices of the two symmetric modes of the standard tri-coupler (excluding the antisymmetric mode), and $\lambda$ is the wavelength.

These effective refractive indices were simulated for a standard tri-coupler with a 2\,\si{\um} edge-to-edge gap and $\lambda =$1.55\,\si{\um}, determining the length. Afterwards, a fine-tuning step of iterative beam propagation simulations was performed to determine the coupling length for maximum exoplanet throughput into the central output port, yielding a length of 308\,\si{\um}.

Simulations at 1.55\,\si{\um} (Fig.\,\ref{fig:BP_Examples}) confirm the expected reciprocal behavior: equal fields at 0° phase difference launched into the outer ports constructively interfere in the center output port. In addition, equal fields with a 180° phase offset will destructively interfere in the center output port. These simulations demonstrate a stable (achromatic) destructive interference throughout the length (see the third column of Fig.\,\ref{fig:BP_Examples}). The maximized exoplanet throughput will exhibit wavelength dependence based on Eq.\,\eqref{eq:BeatLength}. 

\subsubsection{Multimode interference tri-coupler}
\label{sec:MMI}
An alternative broadband approach employs a 3x3~MMI. The behavior of which is calculated using constructive and destructive interference of the multiple ($>20$) excited modes~\cite{Bryngdahl1973}. The positions of self-images along the MMI length~($L$) are given by
\begin{equation}
    L = \frac{M}{N}\frac{4nW^2}{\lambda},
    \label{eq:Combined}
\end{equation}
where $n$ is the core refractive index, $W$ is the MMI width, $N$ is the number of images (for example, $N=2$ to split an input into two images), and $M$ is the iteration of that number of images ($M=1$ for the first iteration and thus the shortest length), where $M$ and $N$ are coprime integers.

Figure\,\ref{fig:BP_Examples} (the second row, third column) shows that when the MMI’s two output ports are injected with equal-intensity electric fields having a 180° phase offset, it exhibits a central null analogous to that of the evanescent tri-couplers. Unlike those devices, however, the MMI’s null depth varies with length as the internal modes alternate between forming two and four images. From Eq.\,\eqref{eq:Combined}, this length dependence implies a wavelength dependence, but, as shown in Sec.\,\ref{sec:FabTol}, this is not a limiting factor.

A 30\,\si{\um}-wide MMI was determined through simulation to be the smallest viable width (without eliciting evanescent coupling between input/output waveguides), and a device length of 440\,\si{\um} produced no light in the central output port when treated as a splitter.

This configuration corresponds to a symmetrical MMI mode~\cite{Soldano1995} that reduces the length to a quarter of what is calculated in Eq.\,\eqref{eq:Combined} (for $M=1$ and $N=2$), hence the 440\,\si{\um} length. Launching into the outer waveguides won't benefit from the reduced MMI length and, as such, corresponds to a general interference MMI for $N=8$, producing eight images. Adjusting the outer port positions enabled coupling into special MMI configurations that yield three images. 

The input position analysis followed the approach of Bachmann~et~al.~\cite{Bachmann1995}. Table\,\ref{tab:MMI_Splittings} summarizes Bachmann’s results, highlighting unconventional three-way splitting ratios suitable for a tri-coupler implementation.

\begin{table}[ht!]
    \centering
    \caption{MMI image formation at specific lengths, from Eq.\,\eqref{eq:Combined}, for different input positions (where $x=0$ is the MMI's center).}
    \begin{tabular}{|c|c|c|}\hline
       $L({M=1},N)$  & Input Position ($x=W/2-iW/N$) & Split Ratio  \\\hline
        $N=6$ & $i=1,5$ & 62:33:5 \\
        $N=6$ & $i=3$ & 33:33:33 \\
        $N=7$ & $i=1,\dots,6$ & 54:35:11 \\
        $N=8$ & $i=2,6$ & 50:25:25 \\\hline
    \end{tabular}
    \label{tab:MMI_Splittings}
\end{table}

Tapered access ports of 120\,\si{\um} length and 5.85\,\si{\um} width were positioned at  $\pm$8.24\,\si{\um} ($x=2W/7.28$: between $N=7$ and $8$ from Tab.\,\ref{tab:MMI_Splittings}), balancing compactness and coupling efficiency with fabrication requirements (a gap between waveguides of 2\,\si{\um}). This position was determined through iterative simulations of the outer waveguide positions, whilst monitoring the three outer outputs for maximal light in the center output.  

\subsubsection{Tapered tri-coupler}
To mitigate exoplanet throughput chromaticity, the outer waveguides were linearly tapered along the coupling region (shown in the third row of Fig.\,\ref{fig:BP_Examples}), following the approach of Hsiao~et~al.~\cite{Hsiao2010} for an infinite-length tapered tri-coupler. In a tapered tri-coupler, the coupling coefficient and the propagation constants vary along the propagation direction \textit{z}. By finely tuning these two through waveguide width selection and the rate of change by controlling the length, it's possible to get similar exoplanet light behavior over a large waveband. To the author's best knowledge, there is no analytical solution for the tapered tri-coupler, though progress has been made for tapered directional couplers~\cite{Milton1975, Smith1976, Takagi1992}. Hence, the tri-coupler parameters were determined through a systematic optimization procedure:

\begin{itemize}
  [ topsep=5pt, itemsep=0pt]

    \item Begin with an initial design with a large $W_L$/$W_S$ contrast and an extended interaction length
    \item Simulate the device to identify the position at which light concentrates in the central waveguide
    \item Set the new design at this interaction length, and adjust $W_L$ accordingly
    \item Perform a broadband simulation of exoplanet throughput to evaluate achromaticity
    \item Incrementally increase $W_S$ and repeat the above steps
    \item Repeat, until no further improvement in exoplanet throughput across the full bandwidth is achieved 
\end{itemize}
This procedure was undertaken for central waveguide widths of 2.8, 3.0, and 3.2\,\si{\um} (a parameter space restricted by simulation time).

The tapered tri-coupler (Fig.\,\ref{fig:CAD_TaperedTricoupler}) used a 450\,\si{\um} coupling length with the central waveguide fixed at 2.8\,\si{\um}, while the outer arms decreased linearly from 4.0\,\si{\um} to 2.0\,\si{\um}. This geometry maintains single-mode operation while achieving partial $L_{\pi}$ achromatization (see Sec.\,\ref{sec:Results}). 

The corresponding simulation (Fig.\,\ref{fig:BP_Examples}) demonstrates the same reciprocal nulling response as the standard tri-coupler. However, because the taper introduces asymmetry between the outer arms, the optimal configuration depends jointly on the taper gradient, coupling length, and central waveguide width. Although longer tapered structures can enhance achromaticity~\cite{Milton1975, Ramadan1998}, the device presented here was designed for compactness, balancing broadband performance with chip-scale integration requirements. Tapered bends were employed at the input and output to connect to 3.4\,\si{\um} access waveguides, introducing minor additional loss in star light (see Sec.\,\ref{sec:Loss}) relative to the standard tri-coupler.

These three tri-coupler designs — the standard evanescent coupler, MMI, and tapered evanescent coupler — were compared for their suitability as exoplanet beam combiners and starlight attenuators using the parameters above.

\section{Results}
\label{sec:Results}
In this section, we compare the simulated performance of the three tri-coupler configurations, examining exoplanet throughput, their null depth, and potential sensitivity of the null performance to fabrication errors. The expected propagation and radiative losses are also analyzed. The potential of each design for fringe tracking is investigated to assess how effectively the null can be maintained under varying phase conditions.

The power at each output port was monitored using a square $10\times10$\,\si{\um} partial power monitor, and normalized by the sum across all output monitors. These monitors were selected to encompass the mode guided by the waveguide plus scattered light in the vicinity that may disturb the signal, which is especially important when discussing the starlight attenuation as will be shown later in Sec.\,\ref{sec:Null}.

\begin{figure}[htbp]
    \centering
    \setlength{\tabcolsep}{1pt}
    \renewcommand{\arraystretch}{1.3}

    \begin{tabular}{@{} c c c c @{}}
        & \hspace{12pt} \textbf{Outer (Left)} & \hspace{8pt}\textbf{Center} & \textbf{Outer (Right)} \\[4pt]

        \multirow{3}{*}[3.4cm]{\centering\rotatebox{90}{\textbf{Standard tri-coupler}}} &
        \begin{subfigure}{0.3\textwidth}
            \centering
            \includegraphics[width=4.2cm,height=3.2cm]{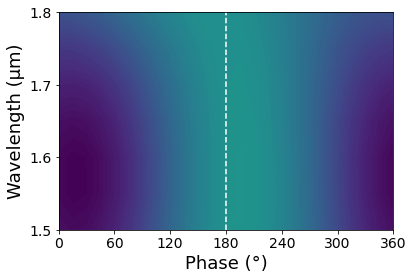}
            \label{fig:Tricoupler_PhaseWvl_Left}
        \end{subfigure} &
        \hspace{8pt}\begin{subfigure}{0.3\textwidth}
            \centering
            \includegraphics[width=\linewidth,height=3.2cm]{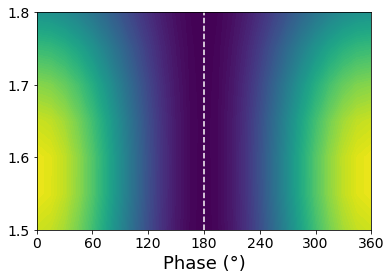}
            \label{fig:Tricoupler_PhaseWvl_center}
        \end{subfigure} &
        \begin{subfigure}{0.3\textwidth}
            \centering
            \includegraphics[width=4.7cm,height=3.2cm]{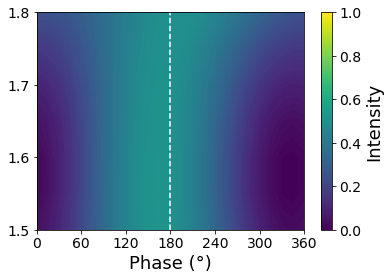}
            \label{fig:Tricoupler_PhaseWvl_Right}
        \end{subfigure} \\[-20pt]
       \multirow{3}{*}[2.6cm]{\centering\vspace{1.5mm}\rotatebox{90}{\textbf{MMI}}} &
        \begin{subfigure}{0.3\textwidth}
            \centering
            \includegraphics[width=4.2cm,height=3.2cm]{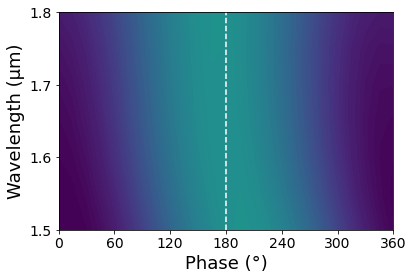}
            \label{fig:MMI_Left_PhaseWvl}
        \end{subfigure} &
        \hspace{8pt}\begin{subfigure}{0.3\textwidth}
            \centering
            \includegraphics[width=\linewidth,height=3.2cm]{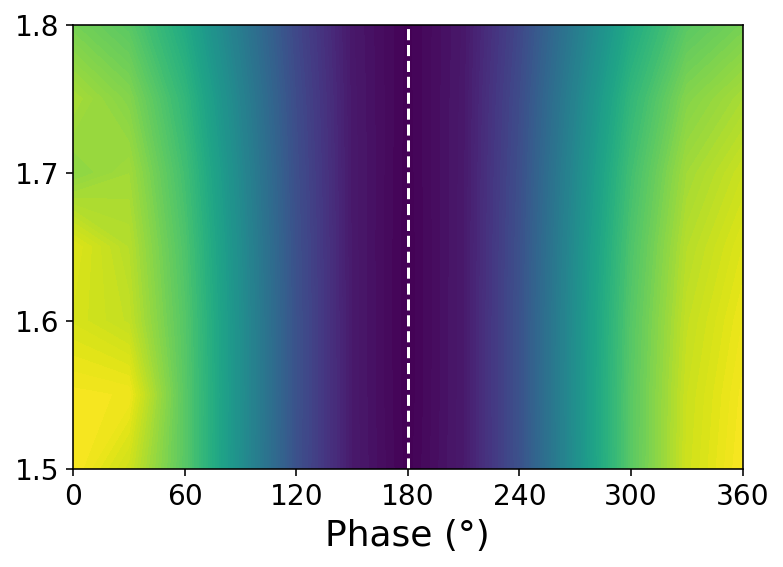}
            \label{fig:MMI_center_PhaseWvl}
        \end{subfigure} &
        \begin{subfigure}{0.3\textwidth}
            \centering
            \includegraphics[width=4.7cm,height=3.2cm]{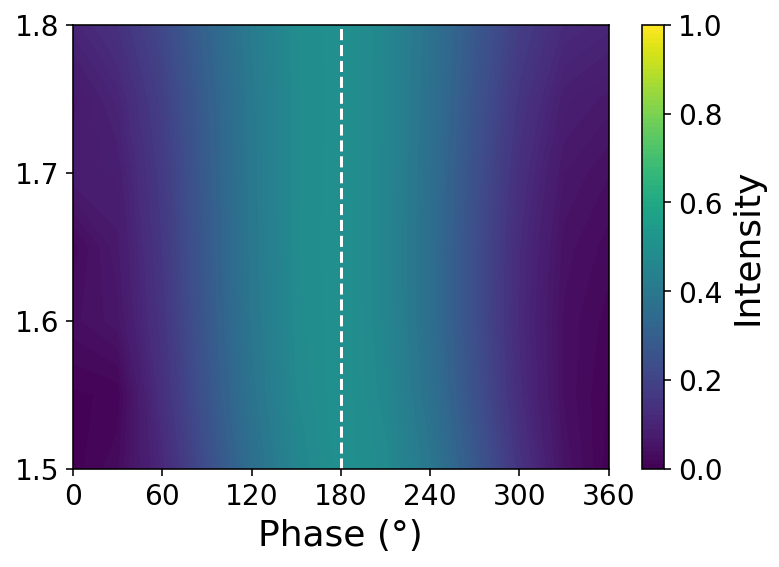}
            \label{fig:MMI_Right_PhaseWvl}
        \end{subfigure} \\[-20pt]
        \multirow{3}{*}[3.4cm]{\centering\rotatebox{90}{\textbf{Tapered tri-coupler}}} &
        \begin{subfigure}{0.3\textwidth}
            \centering
            \includegraphics[width=4.2cm,height=3.2cm]{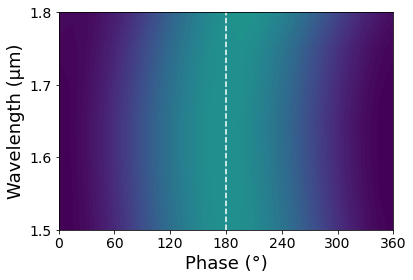}
            \label{fig:TaperedCoupler_Left_PhaseWvl}
        \end{subfigure} &
        \hspace{8pt}\begin{subfigure}{0.3\textwidth}
            \centering
            \includegraphics[width=\linewidth,height=3.2cm]{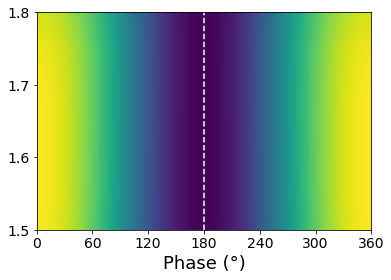}
            \label{fig:TaperedCoupler_center_PhaseWvl}
        \end{subfigure} &
        \begin{subfigure}{0.3\textwidth}
            \centering
            \includegraphics[width=4.7cm,height=3.2cm]{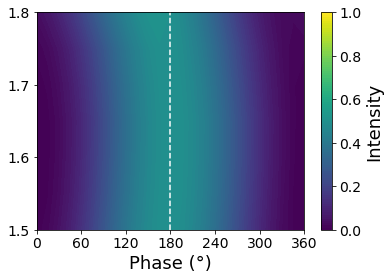}
            \label{fig:TaperedCoupler_Right_PhaseWvl}
        \end{subfigure}\\[-20pt]
    \end{tabular}
    \caption{Normalized intensity in the three tri-coupler configurations (Standard, MMI, and Tapered) for the left, center, and right output ports. Equal-intensity launch fields are injected into the two outer input ports across the full wavelength range as a function of phase difference.}
    \label{fig:PhaseWvl}
\end{figure}

To understand how these tri-couplers perform for nulling, maximizing exoplanet light, and fringe tracking, the light distributions at the output ports were simulated for a range of phase offsets between the two outer input waveguides, across the full operational bandwidth. The resulting output intensities for the outer and central waveguides are shown in Fig.\,\ref{fig:PhaseWvl}, normalized for lossless propagation.

All three devices shown in Fig.\,\ref{fig:PhaseWvl} exhibit similar phase-dependent responses. For the left and right output ports, the responses are mirror images of one another.  It is crucial for differential sensing that these responses do not overlap, discussed in Sec.\,\ref{sec:Sensing}.

In each case, the central output port shows maximum throughput at a 0$^{\circ}$ phase offset and a deep null at 180$^{\circ}$, consistent with the expected reciprocal symmetry. This null corresponds to destructive interference of the starlight, while the 0$^{\circ}$ peak represents the maximum possible exoplanet throughput (for a planet at $\lambda/2D$). Beyond 180$^{\circ}$, the response returns in a symmetric periodic pattern, demonstrating the full interferometric modulation across 360$^{\circ}$.

The tapered tri-coupler appears to have the highest achromatic exoplanet throughput among the other tri-couplers. This is best examined using a cross-section of these images.

\subsection{Exoplanet throughput}
\label{sec:Throughput}

\begin{figure}[ht]
    \centering
    \includegraphics[trim={0 0.7cm 0 0.8cm},width=0.7\linewidth]{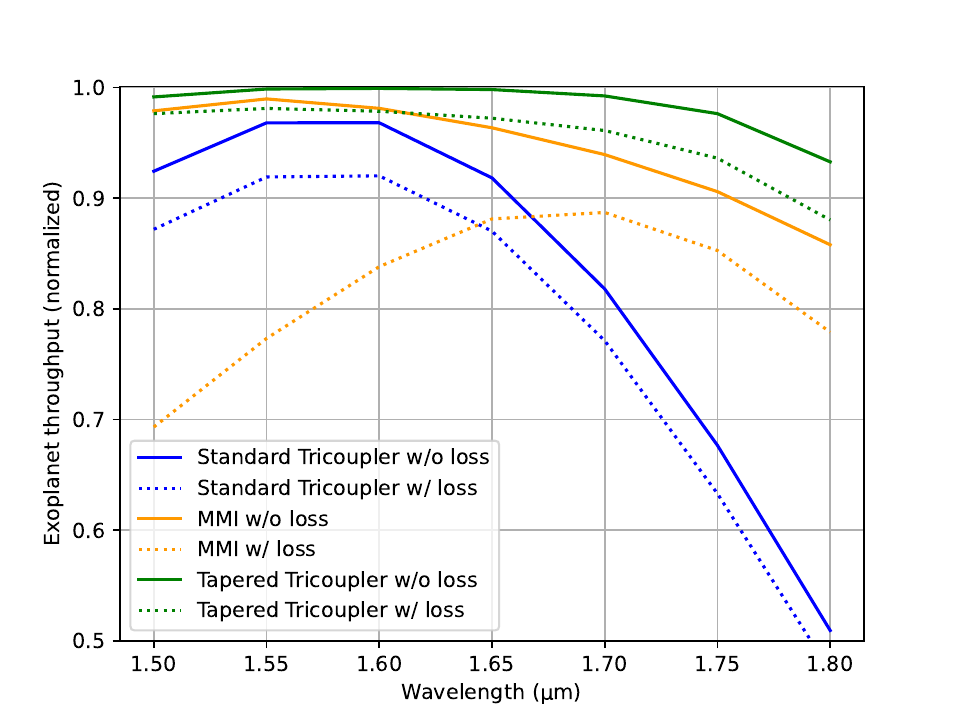}
    \caption{Comparison of constructive interference in the central output when injecting equal-intensity electric fields with a 0° phase difference into the outer waveguides, for the three tri-coupler designs, normalized to total throughput.}
    \label{fig:AntiNull_Comparision}
\end{figure}

Figure\,\ref{fig:AntiNull_Comparision} presents the exoplanet throughput corresponding to the wavelength cross-sections at 0° phase offset of Fig.\,\ref{fig:PhaseWvl}: the light in the central output port, normalized to the light in all three output ports. Figure\,\ref{fig:AntiNull_Comparision} also compares exoplanet throughput without normalization, incorporating the component loss. This highlights the tri-coupler exoplanet throughput ``performance'' and the ``real'' exoplanet throughput to be expected by the coupler. 

The standard and tapered tri-coupler, and the MMI data set without losses, reach peak efficiency near 1.55\,\si{\um}, corresponding to their design wavelength. The standard tri-coupler's throughput drops to $\sim$50\% at 1.8\,\si{\um}. This makes the standard tri-coupler the worst performing of the three, for both datasets with and without losses. The normalized tapered tri-coupler exhibits an almost achromatic response of 100\% throughput, and 97\% when not. Only when loss is included does the throughput drop below 90\%. The MMI, including loss, is the only tri-coupler with a low throughput at 1.55\,\si{\um} due to its high loss away from the design wavelength, potentially due to the outer waveguide position as described in Sec.\,\ref{sec:MMI}. It has a maximum exoplanet throughput of 88\% at 1.7\,\si{\um}. 

These results show that both the MMI and tapered tri-couplers achieve higher exoplanet throughput than the standard design, with the tapered configuration offering the best overall performance.

\subsection{Component and bend losses}
\label{sec:Loss}

The component's loss is a major distinction between evanescent tri-couplers and the MMI design. Evanescent couplers are effectively lossless (0.014\,dB$\approx$ 0.3\% loss calculated for these devices), except for the radiative losses introduced by the bent waveguides that connect the coupling region to the rest of the chip, the subject of this section.

MMIs are known to exhibit higher and more wavelength-dependent losses, particularly when optimized for equal splitting across a broad bandwidth~\cite{KenchingtonGoldsmith2017a, KenchingtonGoldsmith2017b}. This wavelength dependence arises from the beat-length relationship described in Eq.\,\eqref{eq:Combined}, which predicts maximum throughput near the design wavelength (1.55\,\si{\um}) and reduced throughput at shorter and longer wavelengths.

\begin{figure}[ht]
    \centering
    \includegraphics[trim={0 0.7cm 0 0.75cm},width=0.7\linewidth]{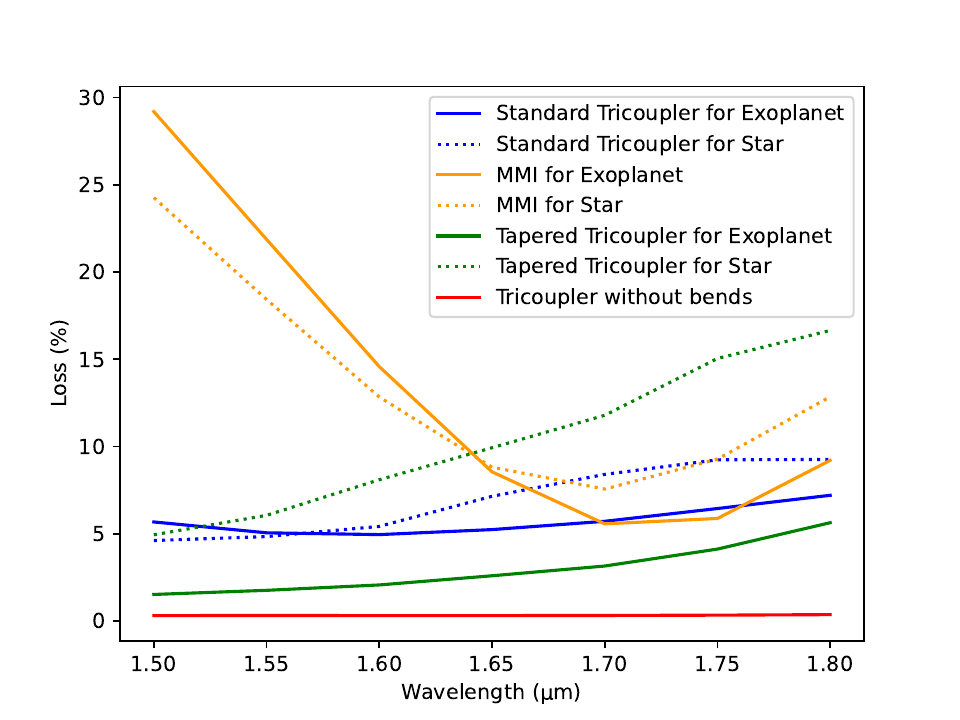}
    \caption{Simulated loss of the MMI (orange) compared with the standard tri-coupler (blue), tapered tri-coupler (green), and a standard tri-coupler without bends (red).}
    \label{fig:Loss}
\end{figure}

Figure\,\ref{fig:Loss} compares the simulated total losses of all three tri-couplers when exoplanet light is passed through the tri-couplers and when starlight is passed through the tri-coupler. It includes the bend losses of each device and assumes no material losses. The MMI displays a narrow region of low loss comparable to the other designs (shifted in wavelength by the custom outer waveguide position), but with a more pronounced wavelength dependence.

It shows that the MMI and standard tri-coupler have a similar loss in both exoplanet and star light, but the tapered tri-coupler has a considerably higher star light loss than exoplanet light loss. This is due to a mode mismatch between the coupler's supermodes and the bent waveguide modes. Increasing the bend radius should reduce this loss for both star and exoplanet light and the difference between the losses. 

The “no-bend” tri-coupler represents an idealized (impossible in reality) case in which parallel waveguides begin and end at 2\,\si{\um} and remain without deviation throughout the simulation (evanescently coupling light throughout the simulation). When bends are introduced, both evanescent designs exhibit increasing loss with wavelength. This trend reflects bend-induced radiation, as larger wavelengths correspond to larger mode sizes and consequently greater leakage in curved sections.

The tapered tri-coupler has slightly higher loss than the standard design, primarily because one set of bent waveguides narrows to smaller widths, which have greater loss than the standard waveguide bends, and another set with larger widths, which may excite higher-order modes that radiate from the bend. The standard-waveguide junction loss for a 0.85\,mm bend radius was calculated in RSoft FemSIM at 3.6\% (0.16\,dB) at 1.55\,\si{\um}. This is higher than what was simulated for exoplanet light in a tapered tri-coupler, reflecting the difference between losses from a finite element simulation and from the beam propagation method reported in Fig.\,\ref{fig:Loss}. 

Comparing the tri-coupler without bends to the other evanescent tri-couplers indicates that the vast majority of loss in the two evanescent tri-couplers is due to bend losses, which can be improved by increasing the bend radius. The MMI, by contrast, exhibits substantially higher intrinsic loss at the shorter wavelengths. This has yet to be corrected by the means discussed in this work. 

\subsection{Starlight attenuation performance and fabrication tolerance}
\label{sec:Null}
\label{sec:FabTol}

The measurable starlight attenuation is limited both intrinsically, by radiative light from bends and the scattered light from the components, specifically the MMI, and extrinsically, by fabrication imperfections such as waveguide asymmetry and under-etching. The following simulations quantify both effects.

The scattered light from starlight injection was simulated as a point-source input equally into the two outer input ports with a 180° phase offset. Ideally, the null depth in the tri-coupler simulations should be limited only by the simulation's numerical precision floor of approximately 80\,dB ($10^{-8}$).

\begin{figure}[ht]
    \centering
    \begin{subfigure}{0.45\textwidth}
            \includegraphics[trim={1.5cm 0.5cm 1.2cm 1.0cm},width=0.95\linewidth]{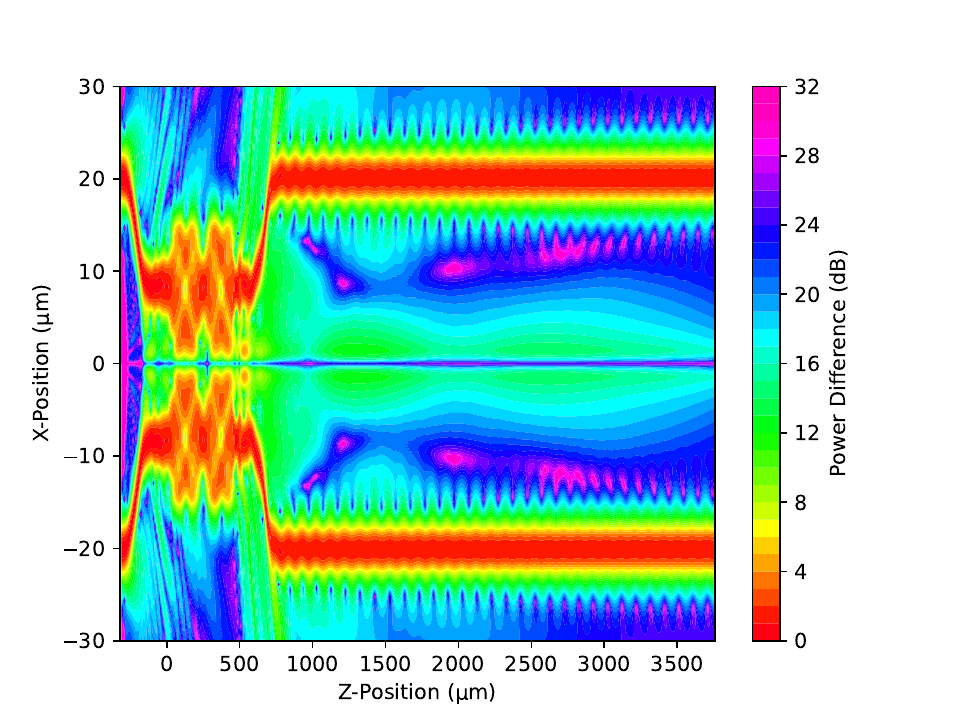}
        \caption{MMI, Small bends}
        \label{fig:ScatteredLight_MMI_0p6}
    \end{subfigure}
    \begin{subfigure}{0.45\textwidth}
            \includegraphics[trim={1.5cm 0.5cm 1.5cm 1.0cm},width=0.95\linewidth]{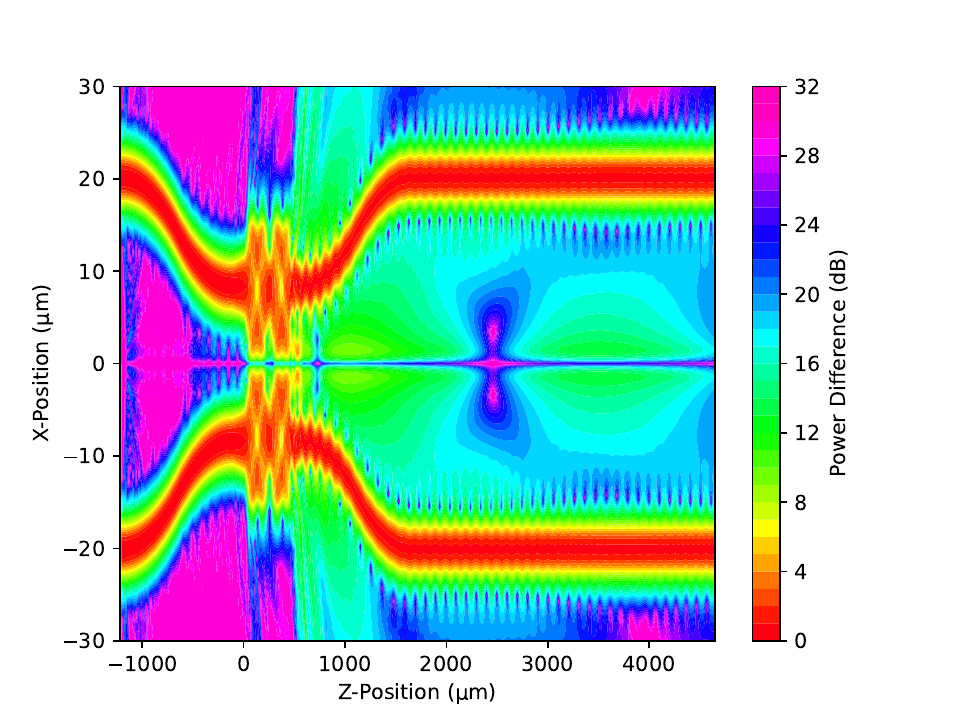}
        \caption{MMI, Large bends}
        \label{fig:ScatteredLight_MMI_20}
    \end{subfigure}
    
       \begin{subfigure}{0.45\textwidth}
            \includegraphics[trim={1.5cm 0.5cm 1.2cm 0.2cm},width=0.95\linewidth]{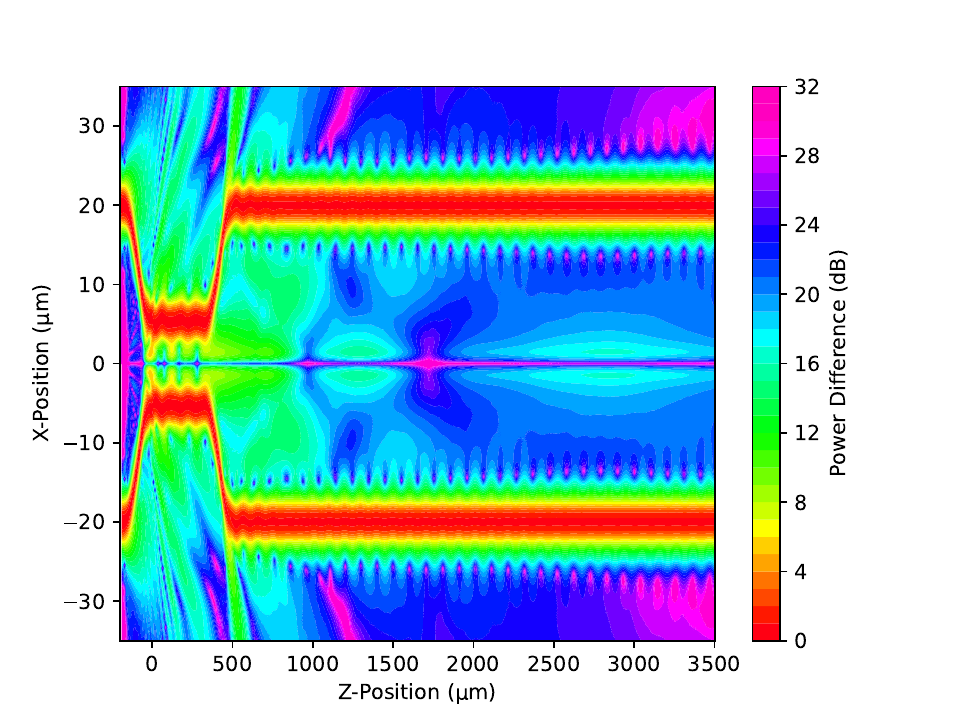}
        \caption{Tri-coupler, Small bends}
        \label{fig:ScatteredLight_tricoupler_0p6}
    \end{subfigure}
    \begin{subfigure}{0.45\textwidth}
            \includegraphics[trim={1.5cm 0.5cm 1.5cm 0.2cm},width=0.95\linewidth]{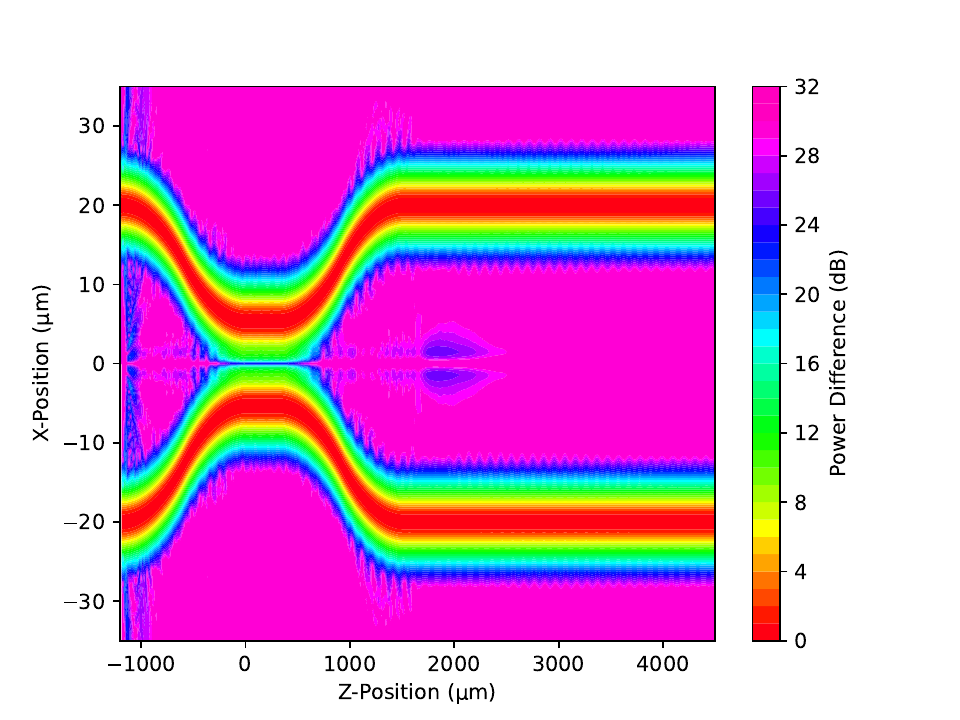}
        \caption{Tri-coupler, Large bends}
        \label{fig:ScatteredLight_tricoupler_20}
    \end{subfigure}
    \caption{Simulated scattered light (normalized) for MMIs and standard tri-couplers for small and large bend radii.}
    \label{fig:ScatteredLight}
\end{figure}

Figure\,\ref{fig:ScatteredLight} presents four scattered light simulations at a 1.55\,\si{\um} wavelength: two standard tri-couplers and two MMIs. The bend radius was  derived using the cosine bend radius equation:
\begin{equation*}
    \text{Bend radius} =\frac{2}{\pi^2}\frac{\text{(Length)}^2}{\text{Width}}
\end{equation*}
for the full \textit{Length} and \textit{Width} of the bent waveguide. The two bend radii included in Fig.\,\ref{fig:ScatteredLight} are to highlight a small bend ($\sim$0.85\,mm) and a large bend ($\sim$20\,mm).
\vspace{0.1cm}

In all cases, except for the standard tri-coupler with a large bend radius (Fig.\,\ref{fig:ScatteredLight_tricoupler_20}), scattered light is evident around the central extinction (at \textit{X-Position} = 0\,\si{\um}), reducing the achievable starlight attenuation, as defined by Eq.\,\eqref{eq:Attenutation}, to approximately 30\,dB~$\left(10^{-3}\right)$.

The standard tri-coupler simulations in Figs.\,\ref{fig:ScatteredLight_tricoupler_0p6} and\,\ref{fig:ScatteredLight_tricoupler_20} demonstrate the effect of bend radii on radiative light, showing a two-order-of-magnitude reduction in scattered light with the bend radius increase --- although there is evidence of scattered light remaining around the center line for the larger bend radius coupler, as shown at \textit{Z-Position} 2000\,\si{\um} for Fig.\,\ref{fig:ScatteredLight_tricoupler_20}, which will reduce the measurable null depth.

While losses of a few percent per bend affect exoplanet throughput weakly, they have a dramatic effect on the ability to attenuate starlight in the central port. Given the expected stellar luminosity relative to exoplanet light, this is critical. Since it's very faint and limiting the null above the numerical simulation floor, careful optimization of bends requires consideration of scattered light pathways and interaction with the nulled output.

For the MMI, radiative scattering from light not coupled into the outer tapers is the dominant factor limiting attenuation. Comparing Figs.\,\ref{fig:ScatteredLight_MMI_20} and\,\ref{fig:ScatteredLight_tricoupler_20}, the scattered light intensity also differs by roughly two orders of magnitude. While further optimization could increase confinement in the MMI, wavelength dependence ensures scattered light will reappear across the band.

To mitigate this effect, the output ports could be physically rerouted using a side-step~\cite{Jovanovic2012}, displacing their output facets away from scattered-light regions. This approach would require careful chip layout to avoid overlap between scattered fields and the output ports.

\subsubsection{Fabrication defects}
While scattered light imposes a fundamental limitation on the achievable attenuation, fabrication imperfections introduce additional degradation, often exceeding the simulated scattering effects. The following analysis quantifies the impact of these fabrication errors on the tri-couplers’ starlight attenuation performance and exoplanet throughput.

\begin{figure}[ht]
\centering
\includegraphics[width=0.8\linewidth]{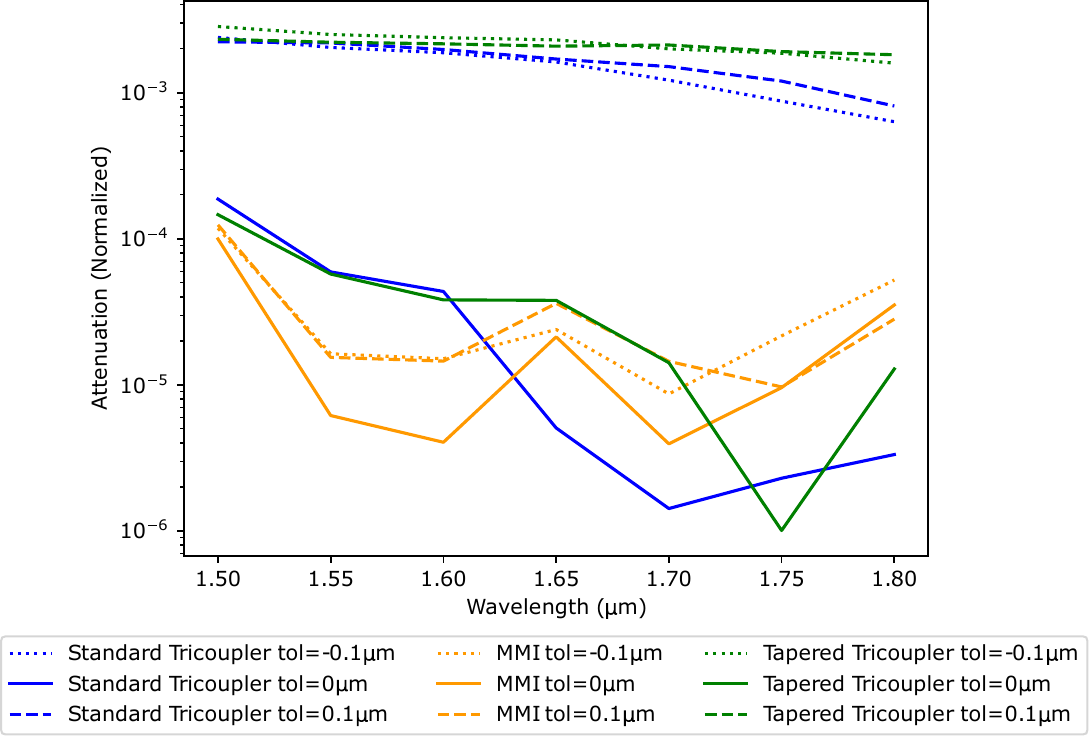}
\caption{Simulated attenuation for each tri-coupler with a $\pm0.1$\,\si{\um} lateral shift of the right-most waveguide.}
\label{fig:OnAxis_Asym_Tol}
\end{figure}

Consider what an asymmetry, due to an imperfection in the fabrication or in the waveguide position, will do to the attenuation. Figure\,\ref{fig:OnAxis_Asym_Tol} compares the attenuation of all three tri-couplers when the right-most waveguide is laterally shifted by $\pm0.1$\,\si{\um}. For the MMI, this shift corresponds to the input position within its broad central region, with the reciprocal output position shifted accordingly.

For the tolerance analysis simulations,  long bent waveguides were used after the tri-coupler's original bends. This allows scattered light to disperse through the simulation for a more accurate representation of the attenuation. For the ideal (defect-free) designs, the attenuation remained limited by scattered light, beginning at 40\,dB~$\left(10^{-4}\right)$ (the MMI) near 1.5\,\si{\um} and improving to between 50\,dB~$\left(10^{-5}\right)$ (the standard tri-coupler) and 60\,dB~$\left(10^{-6}\right)$ (the tapered tri-coupler) at longer wavelengths.

Figure\,\ref{fig:OnAxis_Asym_Tol} shows that the MMI is the least affected by positional asymmetry, remaining below 40\,dB. The standard and tapered tri-coupler degrades to between 20\,dB~$\left(10^{-2}\right)$ and 30\,dB~$\left(10^{-3}\right)$.

The null depth, corresponding to the ratio of the attenuated starlight and the retained exoplanet in the central output port, calculated using Eq.\,\eqref{eq:null}, is 40\,dB over (almost) the entire band without imperfections. Comparatively, the null depth achieved in the tri-couplers used in GLINT was $\sim$30\,dB~ \cite{Martinod2021}, placing this work in a good position if asymmetric fabrication can be avoided. 

\begin{figure}[ht]
    \centering
        \includegraphics[width=0.7\linewidth]{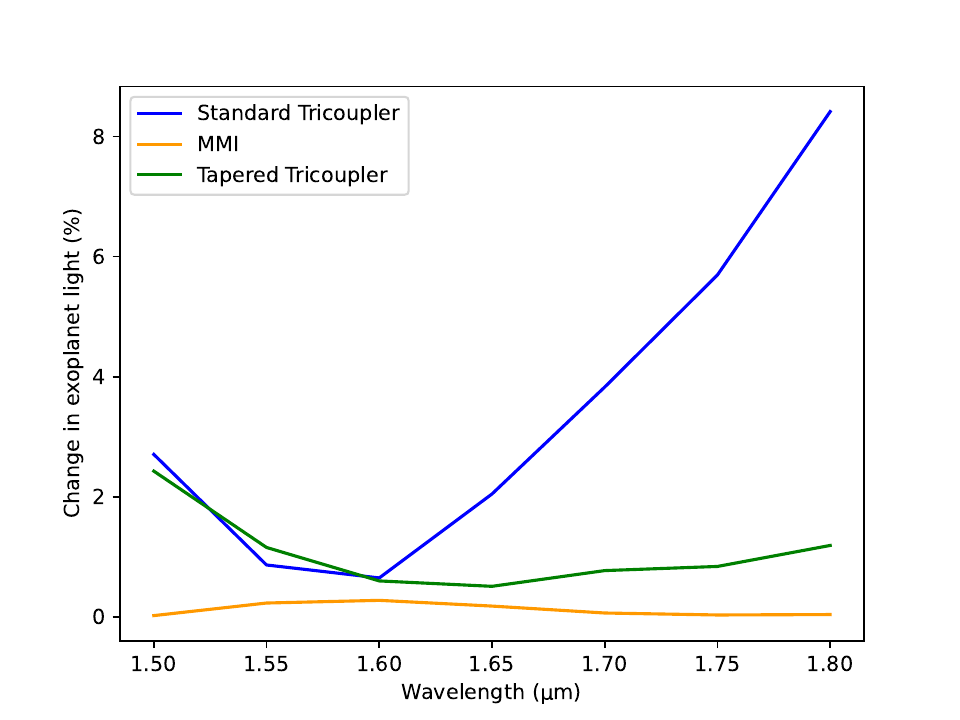}
    \caption{Change in exoplanet light for each tri-coupler when all waveguide widths are reduced by 0.1\,\si{\um}.}
    \label{fig:Symmetric_antinull_tol}
\end{figure}

A more common fabrication issue is a uniform reduction in waveguide width due to under-etching~\cite{Morrissey2015}. This type of error preserves the overall symmetry and therefore the star extinction, but it affects the exoplanet light. Figure\,\ref{fig:Symmetric_antinull_tol} shows the percentage change in exoplanet throughput when all waveguide widths are uniformly reduced by 0.1\,\si{\um}.

The tapered tri-coupler exhibits an average 1\% change in exoplanet light, reaching $>2$\% at the spectral extremes. The standard tri-coupler shows a stronger wavelength dependence, beginning with a 2\% change at short wavelengths and increasing up to an apparent 8\% gain at longer wavelengths. This, for the standard tri-coupler, corresponds to a shift of the constructive interference peak in wavelength --- caused by changes in the beat length resulting from the modified waveguide width and separation.

In contrast, the MMI shows negligible change, consistent with the large multimode region's inherent tolerance. This confirms the MMI’s robustness to common fabrication errors, albeit at the cost of a higher intrinsic loss than the evanescent tri-couplers.

Overall, the achievable attenuation in on-chip tri-couplers is limited by scattered light, bend-induced radiation, and fabrication imperfections. Among the designs studied, the tapered tri-coupler offers the best balance between attenuation and achromaticity while maintaining on-chip fringe tracking, albeit with reduced fabrication tolerance.

\subsection{Fringe tracking}
\label{sec:Sensing}

The rejected starlight exiting the outer ports of the tri-coupler can be used to sense phase errors and drive a control loop to stabilize the null, effectively functioning as an on-chip fringe tracker. This sensing mechanism leverages the differential response of the outer output ports. The phase deviation around the nominal 180° offset, the intensities in the two ports vary in opposite directions, one increases while the other decreases. The sign of this differential change (the slope) indicates the direction of the phase error, while its magnitude reflects the error amplitude. 

Maximum sensitivity is achieved when the intensity response curves of the two outer ports are offset such that their steepest gradients occur near 180°. Conversely, when the two curves overlap and both reach a peak or trough at 180°, the system enters a degenerate state in which no phase discrimination is possible.

\begin{figure}[ht]
    \centering
    \begin{subfigure}{0.45\textwidth}
               \includegraphics[trim={0.9cm 0.6cm 0.9cm 1.0cm},width=\linewidth]{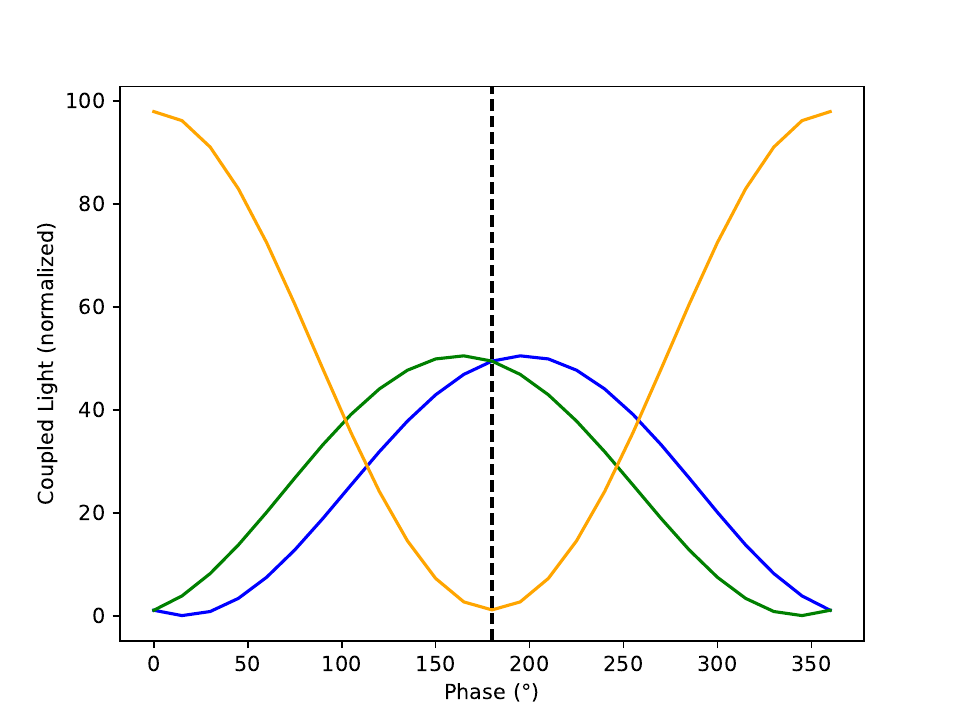}
        \caption{1.5\,\si{\um}}
        \label{fig:MMI_Phase_1p5}
    \end{subfigure}
    \begin{subfigure}{0.45\textwidth}
            \includegraphics[trim={0.9cm 0.6cm 0.9cm 1.0cm},width=\linewidth]{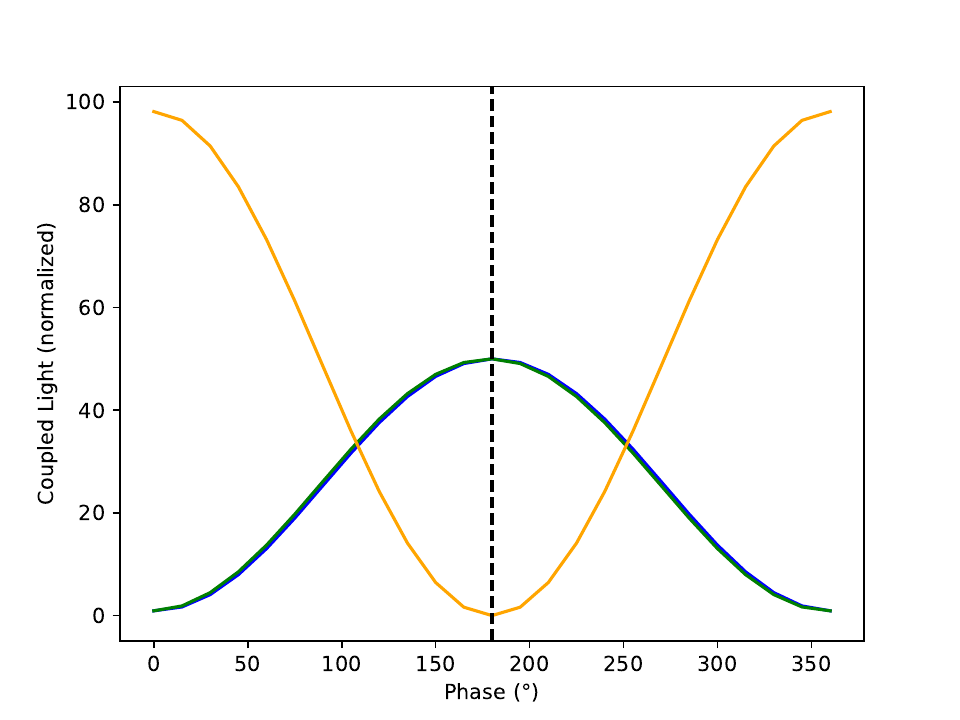}
        \caption{1.6\,\si{\um}}
        \label{fig:MMI_Phase_1p6}
    \end{subfigure}
    \caption{Cross-sections from Fig.\,\ref{fig:PhaseWvl} showing the MMI throughput maps at (a) 1.5\,\si{\um} and (b) 1.6\,\si{\um}. 
    }
    \label{fig:Phase}
\end{figure}

Figure\,\ref{fig:Phase} illustrates two examples of the MMI’s sensing behavior at wavelengths of 1.5\,\si{\um} and 1.6\,\si{\um}. At 1.5\,\si{\um} (Fig.\,\ref{fig:MMI_Phase_1p5}), the outer-port intensities exhibit a clear differential response, allowing effective fringe tracking. In contrast, at 1.6\,\si{\um} (Fig.\,\ref{fig:MMI_Phase_1p6}), the responses are degenerate, making the coupler unsuitable for sensing at that wavelength. This behavior has been observed in prototype ULI tri-couplers that either have degeneracy at 1.4 or 1.55\,\si{\um} when optimized~\cite{Klinner-Teo2022}. 

Multiple non-degenerate wavelengths enable phase-wrap tracking. In addition, the more spectral channels used for fringe tracking, the better the signal-to-noise of the wavefront sensing. The objective is to find a tri-coupler with no degeneracy points over the bandwidth.

\begin{figure}[ht]
    \centering
    \includegraphics[trim={0 0.6cm 0 0.75cm},width=0.7\linewidth]{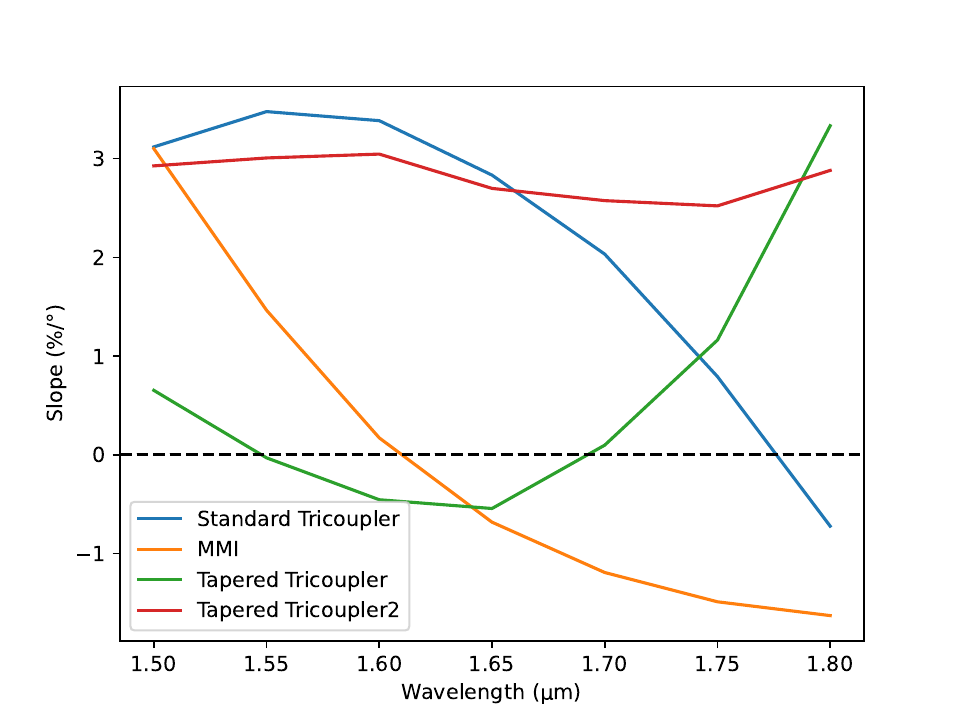}
    \caption{Comparison of the outer-left-port slope at 180° phase offset for the three tri-couplers, including an additional tapered design.}
    \label{fig:LinearityComparision}
\end{figure}

The slope, the derivative, of either outer port’s intensity response at 180° provides a simple metric for identifying degeneracy. Figure\,\ref{fig:LinearityComparision} shows the slope of the left outer port for each tri-coupler design. The right outer port slope will have the same magnitude but opposite sign - due to conservation of energy. The standard tri-coupler and MMI both exhibit large initial slopes, indicating strong sensitivity, before passing through zero, where the outputs become degenerate at that wavelength, and then transitioning to a negative slope, corresponding to a reversal of the left and right port responses relative to Fig.\,\ref{fig:MMI_Phase_1p5}. 

The standard tri-coupler and MMI have a single degeneracy point, whereas the tapered tri-coupler crosses the degeneracy line twice and remains close to 0\,\%/° until the 1.75\,\si{\um} wavelength.

A second tapered tri-coupler, like that in Fig.\,\ref{fig:CAD_TaperedTricoupler} with a 400\,\si{\um} length, a central waveguide of 3\,\si{\um} and tapered outer waveguide from 2 to 5\,\si{\um} was included in Fig.\,\ref{fig:LinearityComparision} to demonstrate a configuration that avoids degeneracy entirely. This design enables continuous fringe tracking across the full band. However, this comes at the expense of exoplanet throughput, which is reduced to 85\% at 1.55\,\si{\um} but does reach 96\% at longer wavelengths (excluding losses), indicating that it is possible to create a low-loss tapered tri-coupler without degeneracy.

Changing the parameters of the other tri-couplers may shift the degeneracy point outside this waveband~\cite{Klinner-Teo2022}, but, due to these couplers' dependence on $L_{\pi}$ from Eq.\,\eqref{eq:BeatLength}~and~\eqref{eq:Combined}, they will not be able to have a slope response like tapered tri-coupler~2 from Fig.\,\ref{fig:LinearityComparision}.

If changing the tapered tri-coupler's outer arm widths and increasing the coupling length is sufficient to avoid degeneracy, at the cost of exoplanet throughput, future work could further optimize the tapered coupler by exploring its full parameter space, potentially regaining much of the lost throughput, but this lies beyond the scope of the present study.

\section{Discussion}

The results discussed in Sec.\,\ref{sec:Results}, and summarized in Tab.\,\ref{tab:Summary}, focused on the exoplanet light, the starlight attenuation, and sensing, all three of which are important for exoplanet detection. 

\begin{table}[ht]
    \centering
    \caption{Summary of significant results at 1.55\,\si{\um}, where +/- refer to the range across the simulated band. Highlighted in red are the worst results between the tri-couplers.}
    \begin{tabular}{|p {5.3cm}|c|c|c|c|}\hline
    Metric $\left(\text{at 1.55\,\si{\um}} ^{+\text{ highest value}}_{-\text{ lowest value}}\right)$ & Standard & MMI & Tapered 1 & Tapered 2*\\\hline
    Exoplanet throughput at $\lambda/2D$ ($\eta_p$: \%) &  \cellcolor[rgb]{1.0,0.5,0.5} $97^{+0}_{-47}$&   $99^{+0}_{-14}$&   $100^{+0}_{-6}$ & $85^{+11}_{-8}$\\
    With loss &  \cellcolor[rgb]{1.0,0.5,0.5} $92^{+0.2}_{-45}$ & \cellcolor[rgb]{1.0,0.5,0.5} $77^{+11}_{-8}$ & $98^{+0}_{-10}$ & $82^{+8}_{-6}$\\\hline
    Starlight attenuation ($\eta_S$: dB)&   $42^{+17}_{-5}$ &   $52^{+2}_{-12}$&   $42^{+18}_{-5}$ & \cellcolor[rgb]{1.0,0.7,0} NA\\
    Bandwidth below 40\,dB (nm) &   270 &   300 &   280 & \cellcolor[rgb]{1.0,0.7,0} NA\\\hline
    $\hookrightarrow$ fabrication deviation (dB)& \cellcolor[rgb]{1.0,0.5,0.5} $27^{+4}_{-1}$ &   $48^{+3}_{-9}$& \cellcolor[rgb]{1.0,0.5,0.5} $27^{+1}_{-1}$ & \cellcolor[rgb]{1.0,0.7,0} NA\\
    Bandwidth below 40\,dB (nm) & \cellcolor[rgb]{1.0,0.5,0.5} 0 &   290 & \cellcolor[rgb]{1.0,0.5,0.5} 0& \cellcolor[rgb]{1.0,0.7,0} NA\\\hline
    Null depth (dB) & $42^{+16}_{-5}$ & $52^{+2}_{-13}$ & $43^{+17}_{-4}$ & \cellcolor[rgb]{1.0,0.7,0} NA \\\hline
    Star Loss (\%) &   $5_{-0.3}^{+2.3}$& \cellcolor[rgb]{1.0,0.5,0.5} $21.6^{+7.8}_{-16}$&   $6.2^{+10.4}_{-1.6}$ &$5.9^{+9.1}_{-0.1}$\\
    Exoplanet loss (\%) & $5_{-0.7}^{+4.2}$ & \cellcolor[rgb]{1.0,0.5,0.5} $18.3^{+5.8}_{-11}$&  $1.7^{+4}_{-0.3}$ &$2.8^{+6.2}_{-1.1}$ \\\hline
    Waveguide shrinkage, exoplanet throughput deviation (\%) & \cellcolor[rgb]{1.0,0.5,0.5} $1^{+7}_{-0.2}$ & $0.2^{+0}_{-0.2}$ & $1^{+1}_{-0.5}$ & \cellcolor[rgb]{1.0,0.7,0}  NA\\\hline 
    Fringe tracking non-degenerecy across the entire band & \cellcolor[rgb]{1.0,0.5,0.5}  no& \cellcolor[rgb]{1.0,0.5,0.5}  no&   \cellcolor[rgb]{1.0,0.5,0.5} no & yes\\\hline
    Scattered light mitigation  &   yes& \cellcolor[rgb]{1.0,0.5,0.5} no&   yes & yes \\
    \hline
    
    \end{tabular}
    \text{*The star attenuation and fabrication robustness were not simulated for this tri-coupler.}
    \label{tab:Summary}
\end{table}

From the standard tri-coupler to the MMI to the tapered tri-coupler, we have seen an increase in achromaticity for exoplanet light throughput. Each tri-coupler has $>90$\% throughput (excluding loss) at 1.55\,\si{\um} except for the second tapered tri-coupler, which was optimized for sensing, not exoplanet throughput. From the table, the standard tri-coupler is the worst-performing tri-coupler. At 1.55\,\si{\um} it achieves a 97\% throughput, but at 1.8\,\si{\um} it drops to 50\% throughput. When accounting for exoplanet loss, the MMI also has low throughput. The MMI exoplanet throughput (including loss) simulation has the lowest maximum throughput of 77\% at 1.55\,\si{\um}. The tapered tri-couplers maintain high throughput regardless of loss with little deviation over the entire bandwidth. 

Mathematically, the starlight attenuation is achromatic. The results in Tab.\,\ref{tab:Summary} quantify this, including scattered light overlapping and potentially coupling into the central ``nulled'' output port. The null depth was included under the starlight attenuation, but has very minor deviations from the starlight attenuation results due to the slight differences in system losses between the exoplanet and star. 

The attenuation retention was tested against a common issue of asymmetry. The tri-couplers had the gap between the center and right waveguides changed by $0.1$\,\si{\um} whilst maintaining the center and left (edge-to-edge) waveguide gap at 2\,\si{\um}. This is equivalent to defects in one input arm (the right waveguide). The tolerance analysis shows that asymmetric gaps between waveguides reduce starlight attenuation significantly in evanescent tri-couplers. The MMI is least affected by asymmetry but is still restricted to 40\,dB $\left(10^{-4}\right)$ by the fabrication error.   

Table\,\ref{tab:Summary} lists the bandwidth for which a $>40$\,dB null is achieved for each tri-coupler. This depth establishes parity between starlight and light from young Jupiter-like exoplanets. Parity between starlight and exoplanet light is not mandatory for detection: a deeper null increases the signal-to-noise ratio as per Eq.\,\eqref{eq:sn}, and 40\,dB is a good first step for a broadband nulling interferometer. Without defect, the three tri-couplers achieve $>40$\,dB attenuation over a $\geq270$\,nm bandwidth. Including defects, only the MMI achieves a starlight attenuation $>40$\,dB. 

The exoplanet throughput is also affected by fabrication limitations. Total waveguide shrinkage results in an 8\% throughput deviation for the standard tri-coupler. This is attributed to a shift in $L_{\pi}$ and is effectively a wavelength shift in the exoplanet throughput. The other tri-couplers are much more robust to shrinkage, with only a 2\% deviation for the tapered tri-coupler and an insignificant amount for the MMI. This is a symmetric change and won't affect the null, like the asymmetric error, and is not investigated in this paper.

For each coupler in this paper, sensing is possible, but not for every wavelength. For the tapered tri-coupler, the sensing is near degenerate (0\%/° slope) most of the waveband, making sensing difficult. This may be because this coupler was optimized for high exoplanet throughput.

For the standard tri-coupler, the sensing response and exoplanet light are linked: the area of best sensing corresponds with the highest exoplanet throughput. The other tri-couplers do not show similar behavior. The MMI's link between the exoplanet light and sensing is not as clear-cut as the standard tri-coupler, and the tapered tri-coupler has the opposite relation between sensing and exoplanet throughput: moving away from a zero slope when the exoplanet light decreases from 100\%. From these examples, we see no clear link between sensing and exoplanet throughput, leaving the possibility for an achromatic tapered tri-coupler in exoplanet throughput and a non-degenerate slope with the correct selection of variables.

The loss simulations showed that bend losses are the predominant cause of star and exoplanet light in the evanescent tri-couplers, whereas the MMI had additional losses. The MMI has the highest loss of the three tri-couplers, with up to 20\%~(0.9\,dB) loss occurring at 1.55\,\si{\um} but decreased to 7\% (0.3\,dB) at 1.7\,\si{\um}. The bend losses are mitigated by increasing the bend radii of the input and output bends. An increased bend radius in the evanescent tri-couplers has the downside of increasing the coupling in the bent waveguides. This is accounted for in the standard tri-coupler by decreasing the coupling length. For the tapered tri-coupler, coupling occurring in the bends will change the achromatic response, either for exoplanet light throughput or sensing optimization, and requires a large parameter scan to re-optimize.

It is unclear why the tapered tri-coupler has more starlight lost than exoplanet light. The issue may come from the supermodes coupling with the fundamental modes of the output ports. Regardless, an increased bend radius should reduce it.

The unique losses associated with the MMI cannot be mitigated through the design parameters discussed in Sec.\,\ref{sec:MMI}. Instead, alternative design strategies and fabrication approaches must be considered. Examples include sub-wavelength gratings~\cite{Cheben2018}, which have previously been used to create ultra-broadband couplers~\cite{Maese-Novo2013}, and inverse-design approaches~\cite{Kojima2021}. These techniques could be leveraged to realize broadband couplers and may be applied in future work to improve both evanescent and MMI-based tri-couplers. Such approaches employ nano-photonic structuring to maintain consistent optical behavior across the waveband. Furthermore, with the advent of neural-network-based optimization, it is now possible to generate either fully “black-box” devices~\cite{Piggott2015} or modified variants of existing coupler geometries~\cite{Chen2023} to produce custom nulling interferometers. While these methods are beyond the scope of the present work, they represent promising directions for future development.

\section{Conclusion}

High-contrast imaging enables the detection and characterization of exoplanets far fainter than their host stars. This is typically achieved in interferometry by using multiple telescopes or by using segments of a single telescope, so that stellar light is attenuated while the residual signal is measured. This work has demonstrated that three different tri-couplers can achieve achromatic starlight attenuation and exoplanet throughput using a two-dimensional photonic platform.

Typically, a standard evanescent coupler is used in a nulling interferometry beam combiner. These should be replaced with either an 3X3~MMI or a tapered evanescent tri-coupler. Both can be optimized to increase exoplanet throughput without compromising on the starlight attenuation.


A novel MMI design was proposed that combines the symmetric MMI mode with general MMI interference between $N=7$ and $8$, utilizing injection positions of $\pm$8.24\,\si{\um} for a 30\,\si{\um} wide MMI (a position to MMI width ratio of $\sim$0.27:1) for MMI taper widths of 5.85\,\si{\um} (a taper width to MMI width ratio 0.20:1). This reduces the MMI length to a scale comparable to the tri-coupler. However, it introduces additional losses compared to evanescent couplers, which are limited by bend losses. These additional MMI losses, combined with radiative bending loss, limit exoplanet detection due to starlight scattered into the destructive interference output ports. This necessitates careful routing of the output waveguides to mitigate this noise in the exoplanet detection.

The tapered tri-coupler was optimized for exoplanet throughput across the 1.5--1.8\,\si{\um} waveband. Unlike typical broadband tapered devices that rely on strict adiabaticity, this design, as a splitter, uses a compact design and exploits an extended coupling region into the outer arms.

In terms of sensing, none of the couplers provides perfectly achromatic performance. However, an example tapered coupler demonstrated an achromatic, high-sensing response across the full waveband, at the expense of some exoplanet throughput. This suggests that designs may exist that achieve both high throughput (though not necessarily 100\%) and strong sensing capability.

Each tri-coupler explored in this work achieved a starlight attenuation of $>40$\,dB for $\geq270$\,nm. With the evanescent couplers reaching a null depth of $\sim$60\,dB. The MMI, however, was the only fabricationally robust design, maintaining $>40$\,dB suppression, whereas the evanescent tri-couplers reduced to between 20 to 30\,dB over the band. This makes the MMI the preferred component for assessing the science case of a high contrast imaging instrument until tapered tri-couplers can be fabricated with high precision.

Future work will focus on experimental verification of these components. It will be used to assess fabrication performance, measure starlight attenuation and exoplanet throughput, and determine whether they can be used for sensing. More simulations are required to evaluate the tapered tri-coupler's potential for fringe tracking and establish whether there is a link between exoplanet throughput and degeneracy across the band. And answer the question of whether there is a fundamental trade-off. 

Beyond geometric optimization, incorporating inverse-designed photonic structures offers a promising route toward achieving intrinsically broadband behavior, representing a natural next stage for this work. 

\section{Acknowledgments}
This work was supported by the National Aeronautics and Space Administration~(NASA) under grant number 80NSSC24K1559, awarded through the Astrophysics Research and Analysis~(APRA) program. This work was supported by the National Science Foundation under Grant numbers 2308360 and 2308361. 

\section{Disclosures}
The authors declare no conflicts of interest.


\bibliography{ref,library}

@article{Jovanovic2012,
   author = {N. Jovanovic and P. G. Tuthill and B. Norris and S. Gross and P. Stewart and N. Charles and S. Lacour and M. Ams and J. S. Lawrence and A. Lehmann and C. Niel and J. G. Robertson and G. D. Marshall and M. Ireland and A. Fuerbach and M. J. Withford},
   doi = {10.1111/j.1365-2966.2012.21997.x},
   issn = {00358711},
   issue = {1},
   journal = {Monthly Notices of the Royal Astronomical Society},
   month = {11},
   pages = {806-815},
   title = {Starlight demonstration of the Dragonfly instrument: an integrated photonic pupil-remapping interferometer for high-contrast imaging},
   volume = {427},
   url = {https://academic.oup.com/mnras/article-lookup/doi/10.1111/j.1365-2966.2012.21997.x},
   year = {2012},
}

@article{Hsiao2010,
   author = {Hsien-kai Hsiao and Kim A. Winick and John D. Monnier},
   doi = {10.1364/AO.49.006675},
   isbn = {0003-6935},
   issn = {0003-6935},
   issue = {35},
   journal = {Applied Optics},
   month = {12},
   pages = {6675},
   pmid = {21151223},
   title = {Midinfrared broadband achromatic astronomical beam combiner for nulling interferometry},
   volume = {49},
   url = {https://opg.optica.org/abstract.cfm?URI=ao-49-35-6675},
   year = {2010}
}

@article{Hsiao2009,
   author = {Hsien-kai Hsiao and K. A. Winick and John D. Monnier and Jean-Philippe Berger},
   doi = {10.1364/OE.17.018489},
   isbn = {1094-4087 (Electronic)\r1094-4087 (Linking)},
   issn = {1094-4087},
   issue = {21},
   journal = {Optics Express},
   month = {10},
 
   pages = {18489},
   pmid = {20372579},
   title = {An infrared integrated optic astronomical beam combiner for stellar interferometry at 3-4 microns},
   volume = {17},
   url = {https://opg.optica.org/oe/abstract.cfm?uri=oe-17-21-18489},
   year = {2009}
}

@inproceedings{Gretzinger2019,
   author = {T. Gretzinger and S. Gross and A. Arriola and T. T. Fernandez and D. Strixner and J. Tepper and L. Labadie and M. J. Withford},
   doi = {10.1109/CLEOE-EQEC.2019.8872735},
   isbn = {978-1-7281-0469-0},
   booktitle = {2019 Conference on Lasers and Electro-Optics Europe \& European Quantum Electronics Conference (CLEO/Europe-EQEC)},
   month = {6},
   pages = {1-1},
   publisher = {IEEE},
   title = {Broadband Mid-Infrared Directional and Multimode Interference Couplers in GLS Glass Fabricated using Femtosecond Laser Direct-Writing},
   url = {https://ieeexplore.ieee.org/document/8872735/},
   year = {2019}
}

@article{Bracewell1978,
   author = {R. N. Bracewell},
   doi = {10.1038/274780a0},
   issn = {0028-0836},
   journal = {Nature},
   pages = {780-781},
   title = {Detecting nonsolar planets by spinning infrared interferometer},
   volume = {274},
   year = {1978},
}

@article{Soldano1995,
   author = {Lucas B. Soldano and Erik C M Pennings},
   doi = {10.1109/50.372474},
   issn = {07338724},
   issue = {4},
   journal = {Journal of Lightwave Technology},
   pages = {615-627},
   title = {Optical multi-mode interference devices based on self-imaging: principles and applications},
   volume = {13},
   year = {1995},
}

@article{Norris2014,
   author = {Barnaby Norris and Nick Cvetojevic and Simon Gross and Nemanja Jovanovic and Paul N. Stewart and Ned Charles and Jon S. Lawrence and Michael J. Withford and Peter Tuthill},
   doi = {10.1364/OE.22.018335},
   isbn = {9780819496140},
   issn = {1094-4087},
   issue = {15},
   journal = {Optics Express},
   pages = {18335-53},
   title = {High-performance 3D waveguide architecture for astronomical pupil-remapping interferometry},
   volume = {22},
   url = {http://www.opticsexpress.org/abstract.cfm?URI=oe-22-15-18335},
   year = {2014},
}

@inproceedings{Gillessen2010,
 author = {S. Gillessen and F. Eisenhauer and G. Perrin and W. Brandner and C. Straubmeier and K. Perraut and A. Amorim and M. Sch{\"o}ller and C. Araujo-Hauck and H. Bartko and H. Baumeister and J.-P. Berger and P. Carvas and F. Cassaing and F. Chapron and E. Choquet and Y. Clenet and C. Collin and A. Eckart and P. Fedou and S. Fischer and E. Gendron and R. Genzel and P. Gitton and F. Gonte and A. Gr{\"a}ter and P. Haguenauer and M. Haug and X. Haubois and T. Henning and S. Hippler and R. Hofmann and L. Jocou and S. Kellner and P. Kervella and R. Klein and N. Kudryavtseva and S. Lacour and V. Lapeyrere and W. Laun and P. Lena and R. Lenzen and J. Lima and D. Moratschke and D. Moch and T. Moulin and V. Naranjo and U. Neumann and A. Nolot and T. Paumard and O. Pfuhl and S. Rabien and J. Ramos and J. M. Rees and R.-R. Rohloff and D. Rouan and G. Rousset and A. Sevin and M. Thiel and K. Wagner and M. Wiest and S. Yazici and D. Ziegler},
title = {{GRAVITY: a four telescope beam combiner instrument for the VLTI}},
volume = {7734},
booktitle = {Optical and Infrared Interferometry II},
publisher = {SPIE},
pages = {77340Y},
year = {2010},
doi = {10.1117/12.856689},
URL = {https://doi.org/10.1117/12.856689}
}

@article{Wolszczan1992,
   author = {A. Wolszczan and D. A. Frail},
   doi = {10.1038/355145a0},
   issn = {0028-0836},
   issue = {6356},
   journal = {Nature},
   pages = {145-147},
   title = {A planetary system around the millisecond pulsar \text{PSR1257 + 12}},
   volume = {355},
   year = {1992},
}

@article{Bachmann1995,
   author = {M. Bachmann and P. A. Besse and H. Melchior},
   doi = {10.1364/AO.34.006898},
   issn = {0003-6935},
   issue = {30},
   journal = {Applied Optics},
   pages = {6898-910},
   pmid = {21060551},
   title = {Overlapping-image multimode interference couplers with a reduced number of self-images for uniform and nonuniform power splitting.},
   volume = {34},
   url = {http://www.ncbi.nlm.nih.gov/pubmed/21060551},
   year = {1995},
}

@article{Bryngdahl1973,
   author = {Olof Bryngdahl},
   doi = {10.1364/JOSA.63.000416},
   issn = {0030-3941},
   issue = {4},
   journal = {Journal of the Optical Society of America},
   pages = {416-419},
   title = {Image formation using self-imaging techniques},
   volume = {63},
   year = {1973},
}

@inproceedings{Besse1994,
   author = {Pierre A. Besse and Maurus Bachmann and H. Melchior},
   city = {Florence},
   booktitle = {20th European Conference on Optical Communication},
   pages = {669-672},
   title = {New 1x2 \text{MMI} with Free Selection of Power Splitting Ratios},
   year = {1994},
}

@article{Errmann2015,
   author = {Ronny Errmann and Stefano Minardi and Lucas Labadie and Felix Dreisow and Stefan Nolte and Thomas Pertsch},
   doi = {10.1117/12.2055473},
   isbn = {9780819496140},
   issn = {1996756X},
   issue = {24},
   journal = {Applied Optics},
   pages = {7449-7454},
   title = {Integrated optics interferometric four telescopes nuller},
   volume = {54},
   year = {2015},
}

@article{Setterholm2023,
author = {Benjamin R. Setterholm and John D. Monnier and Jean-Baptiste Le Bouquin and Narsireddy Anugu and Jacob Ennis and Laurent Jocou and Nour Ibrahim and Stefan Kraus and Matthew D. Anderson and Sorabh Chhabra and Isabelle Codron and Christopher D. Farrington and Becky Flores and Tyler Gardner and Mayra Gutierrez and Cyprien Lanthermann and Olli W. Majoinen and Daniel J. Mortimer and Gail H. Schaefer and Nicholas J. Scott and Theo A. ten Brummelaar and Norman  L. Vargas},
title = {{MYSTIC: a high angular resolution K-band imager at CHARA}},
volume = {9},
journal = {Journal of Astronomical Telescopes, Instruments, and Systems},
number = {2},
publisher = {SPIE},
pages = {025006},
year = {2023},
doi = {10.1117/1.JATIS.9.2.025006},
URL = {https://doi.org/10.1117/1.JATIS.9.2.025006}
}

@inproceedings{KenchingtonGoldsmith2016,
   author = {H.-D. {Kenchington Goldsmith} and N. Cvetojevic and M. Ireland and P. Ma and P. Tuthill and B. Eggleton and J. S. Lawrence and S. Debbarma and B. Luther-Davies and S. J. Madden},
   doi = {10.1117/12.2232199},
   journal = {Proc. of SPIE},
   month = {8},
   pages = {990730},
   title = {Chalcogenide glass planar \text{MIR} couplers for future chip based Bracewell interferometers},
 publisher = {SPIE},
booktitle = {Optical and Infrared Interferometry and Imaging V},
   volume = {9907},
   url = {http://proceedings.spiedigitallibrary.org/proceeding.aspx?doi=10.1117/12.2232199},
   year = {2016},
}

@article{KenchingtonGoldsmith2017a,
   author = {Harry-Dean {Kenchington Goldsmith} and Nick Cvetojevic and Michael Ireland and Stephen Madden},
   doi = {10.1364/OE.25.003038},
   issn = {1094-4087},
   issue = {4},
   journal = {Optics Express},
   month = {2},
   pages = {3038},
   title = {Fabrication tolerant chalcogenide mid-infrared multimode interference coupler design with applications for Bracewell nulling interferometry},
   volume = {25},
   url = {https://www.osapublishing.org/abstract.cfm?URI=oe-25-4-3038},
   year = {2017},
}

@article{Gretzinger2015,
   author = {Thomas Gretzinger and Simon Gross and Martin Ams and Alexander Arriola and Michael J. Withford},
   doi = {10.1364/OME.5.002862},
   issn = {2159-3930},
   issue = {12},
   journal = {Opt. Mater. Express},
   pages = {2862-2877},
   title = {Ultrafast laser inscription in chalcogenide glass: thermal versus athermal fabrication},
   volume = {5},
   url = {http://www.osapublishing.org/ome/abstract.cfm?URI=ome-5-12-2862%5Cnhttp://www.osapublishing.org/viewmedia.cfm?uri=ome-5-12-2862&seq=0&html=true},
   year = {2015},
}

@article{KenchingtonGoldsmith2017b,
   author = {Harry-Dean {Kenchington Goldsmith} and Michael Ireland and Pan Ma and Nick Cvetojevic and Stephen Madden},
   doi = {10.1364/OE.25.016813},
   issn = {1094-4087},
   issue = {14},
   journal = {Optics Express},
   pages = {16813-16824},
   title = {Improving the extinction bandwidth of MMI chalcogenide photonic chip based MIR nulling interferometers},
   volume = {25},
   url = {https://www.osapublishing.org/abstract.cfm?URI=oe-25-14-16813},
   year = {2017},
}

@inproceedings{KenchingtonGoldsmith2018,
   author = {Harry-Dean {Kenchington Goldsmith} and Michael Ireland and Pan Ma and Rongping Wang and Barry Luther-Davies and Stephen Madden and Barnaby Norris and Peter Tuthill},
   doi = {10.1117/12.2311904},
   isbn = {9781510619555},
   issn = {1996756X},
   issue = {July},
   booktitle = {Optical and Infrared Interferometry and Imaging VI},
   month = {7},
   pages = {33},
   publisher = {SPIE},
volume = {10701},
   title = {Photonic mid-infrared nulling for exoplanet detection on a planar chalcogenide platform},
   url = {https://www.spiedigitallibrary.org/conference-proceedings-of-spie/10701/2311904/Photonic-mid-infrared-nulling-for-exoplanet-detection-on-a-planar/10.1117/12.2311904.full},
   year = {2018},
}

@article{Morley2016,
   author = {Caroline V. Morley and Heather Knutson and Michael Line and Jonathan J. Fortney and Daniel Thorngren and Mark S. Marley and Dillon Teal and Roxana Lupu},
   doi = {10.3847/1538-3881/153/2/86},
   isbn = {0201721481},
   issn = {1538-3881},
   issue = {2},
   journal = {The Astronomical Journal},
   month = {1},
   pages = {86},
   pmid = {21301655},
   publisher = {IOP Publishing},
   title = {FORWARD AND INVERSE MODELING OF THE EMISSION AND TRANSMISSION SPECTRUM OF GJ 436B: INVESTIGATING METAL ENRICHMENT, TIDAL HEATING, AND CLOUDS},
   volume = {153},
   url = {http://arxiv.org/abs/1610.07632%0Ahttp://dx.doi.org/10.3847/1538-3881/153/2/86 http://stacks.iop.org/1538-3881/153/i=2/a=86?key=crossref.200835e96567713b5a25ec6c2c869380},
   year = {2017},
}

@article{Marois2006,
   author = {Christian Marois and David Lafreniere and Rene Doyon and Bruce Macintosh and Daniel Nadeau},
   doi = {10.1086/500401},
   issn = {0004-637X},
   issue = {1},
   journal = {The Astrophysical Journal},
   month = {4},
   pages = {556-564},
   title = {Angular Differential Imaging: A Powerful High‐Contrast Imaging Technique},
   volume = {641},
   url = {http://arxiv.org/abs/astro-ph/0512335%0Ahttp://dx.doi.org/10.1086/500401 http://stacks.iop.org/0004-637X/641/i=1/a=556},
   year = {2006},
}

@article{Morrissey2015,
   author = {P.E. Morrissey and H. Yang and R.N. Sheehan and B. Corbett and F.H. Peters},
   doi = {10.1016/j.optcom.2014.11.083},
   issn = {00304018},
   journal = {Optics Communications},
   month = {4},
   pages = {26-32},
   publisher = {Elsevier},
   title = {Design and fabrication tolerance analysis of multimode interference couplers},
   volume = {340},
   url = {http://dx.doi.org/10.1016/j.optcom.2014.11.083 https://linkinghub.elsevier.com/retrieve/pii/S0030401814011225},
   year = {2015},
}

@article{Guyon2006,
   author = {O. Guyon and E. A. Pluzhnik and M. J. Kuchner and B. Collins and S. T. Ridgway},
   doi = {10.1086/507630},
   issn = {0067-0049},
   issue = {1},
   journal = {The Astrophysical Journal Supplement Series},
   month = {11},
   pages = {81-99},
   title = {Theoretical Limits on Extrasolar Terrestrial Planet Detection with Coronagraphs},
   volume = {167},
   url = {http://arxiv.org/abs/astro-ph/0608506%0Ahttp://dx.doi.org/10.1086/507630 http://stacks.iop.org/0067-0049/167/i=1/a=81},
   year = {2006},
}

@article{Maese-Novo2013,
   author = {A. Maese-Novo and Robert Halir and S. Romero-García and D. Pérez-Galacho and L. Zavargo-Peche and Alejandro Ortega-Moñux and I. Molina-Fernández and J. G. Wangüemert-Pérez and Pavel Cheben},
   doi = {10.1364/OE.21.007033},
   issn = {1094-4087},
   issue = {6},
   journal = {Optics Express},
   month = {3},
   pages = {7033},
   title = {Wavelength independent multimode interference coupler},
   volume = {21},
   url = {http://doi.wiley.com/10.1002/lpor.201600213 https://www.osapublishing.org/oe/abstract.cfm?uri=oe-21-6-7033},
   year = {2013},
}

@article{Bland-Hawthorn2009,
   author = {Joss Bland-Hawthorn and Pierre Kern},
   doi = {10.1364/OE.17.001880},
   issn = {1094-4087},
   issue = {3},
   journal = {Optics Express},
   month = {2},
   pages = {1880},
   title = {Astrophotonics: a new era for astronomical instruments},
   volume = {17},
   url = {https://www.osapublishing.org/oe/abstract.cfm?uri=oe-17-3-1880},
   year = {2009},
}

@article{Milton1975,
   author = {A F Milton and W K Burns},
   doi = {10.1364/AO.14.001207},
   issn = {0003-6935},
   issue = {5},
   journal = {Applied Optics},
   pages = {1207},
   title = {Tapered Velocity Couplers for Integrated Optics: Design},
   volume = {14},
   url = {https://www.osapublishing.org/abstract.cfm?URI=ao-14-5-1207},
   year = {1975},
}

@article{Takagi1992,
   author = {A. Takagi and K. Jinguji and M. Kawachi},
   doi = {10.1109/50.143072},
   issn = {07338724},
   issue = {6},
   journal = {Journal of Lightwave Technology},
   month = {6},
   pages = {735-746},
   title = {Wavelength characteristics of (2 X 2) optical channel-type directional couplers with symmetric or nonsymmetric coupling structures},
   volume = {10},
   url = {http://ieeexplore.ieee.org/document/143072/},
   year = {1992},
}

@article{Lyot1939,
   author = {B Lyot},
   journal = {Monthly Notices of the Royal Astronomical Society},
   pages = {580},
   title = {A study of the solar corona and prominences without eclipses},
   volume = {99},
   year = {1939},
}

@article{Perrin2006,
   author = {G. Perrin and S. Lacour and J. Woillez and E. Thiebaut},
   doi = {10.1111/j.1365-2966.2006.11063.x},
   issn = {0035-8711},
   issue = {2},
   journal = {Monthly Notices of the Royal Astronomical Society},
   pages = {747-751},
   title = {High dynamic range imaging by pupil single-mode filtering and remapping},
   volume = {373},
   url = {https://academic.oup.com/mnras/article-lookup/doi/10.1111/j.1365-2966.2006.11063.x},
   year = {2006},
}

@article{Smith1976,
   author = {Robert B. Smith},
journal = {J. Opt. Soc. Am.},
number = {9},
pages = {882--892},
publisher = {Optica Publishing Group},
title = {Analytic solutions for linearly tapered directional couplers},
volume = {66},
month = {Sep},
year = {1976},
url = {https://opg.optica.org/abstract.cfm?URI=josa-66-9-882},
doi = {10.1364/JOSA.66.000882},
}

@misc{Ramadan1998,
   author = {Tarek A Ramadan and Robert Scarmozzino and Richard M Osgood},
   isbn = {07338724/98$10.0},
   issue = {2},
   journal = {JOURNAL OF LIGHTWAVE TECHNOLOGY},
   pages = {277},
   title = {Adiabatic Couplers: Design Rules and Optimization},
   volume = {16},
   year = {1998},
}

@article{Jovanovic2023,
   author = {Nemanja Jovanovic and Pradip Gatkine and Narsireddy Anugu and Rodrigo Amezcua-Correa and Ritoban Basu Thakur and Charles Beichman and Chad F. Bender and Jean-Philippe Berger and Azzurra Bigioli and Joss Bland-Hawthorn and Guillaume Bourdarot and Charles M Bradford and Ronald Broeke and Julia Bryant and Kevin Bundy and Ross Cheriton and Nick Cvetojevic and Momen Diab and Scott A Diddams and Aline N Dinkelaker and Jeroen Duis and Stephen Eikenberry and Simon Ellis and Akira Endo and Donald F Figer and Michael P. Fitzgerald and Itandehui Gris-Sanchez and Simon Gross and Ludovic Grossard and Olivier Guyon and Sebastiaan Y Haffert and Samuel Halverson and Robert J Harris and Jinping He and Tobias Herr and Philipp Hottinger and Elsa Huby and Michael Ireland and Rebecca Jenson-Clem and Jeffrey Jewell and Laurent Jocou and Stefan Kraus and Lucas Labadie and Sylvestre Lacour and Romain Laugier and Katarzyna Ławniczuk and Jonathan Lin and Stephanie Leifer and Sergio Leon-Saval and Guillermo Martin and Frantz Martinache and Marc-Antoine Martinod and Benjamin A Mazin and Stefano Minardi and John D Monnier and Reinan Moreira and Denis Mourard and Abani Shankar Nayak and Barnaby Norris and Ewelina Obrzud and Karine Perraut and François Reynaud and Steph Sallum and David Schiminovich and Christian Schwab and Eugene Serbayn and Sherif Soliman and Andreas Stoll and Liang Tang and Peter Tuthill and Kerry Vahala and Gautam Vasisht and Sylvain Veilleux and Alexander B Walter and Edward J Wollack and Yinzi Xin and Zongyin Yang and Stephanos Yerolatsitis and Yang Zhang and Chang-Ling Zou},
   doi = {10.1088/2515-7647/ace869},
   issn = {2515-7647},
   issue = {4},
   journal = {Journal of Physics Photonics},
   month = {10},
   pages = {042501},
   publisher = {Institute of Physics},
   title = {2023 Astrophotonics Roadmap: pathways to realizing multi-functional integrated astrophotonic instruments},
   volume = {5},
   url = {https://iopscience.iop.org/article/10.1088/2515-7647/ace869},
   year = {2023}
}

@article{Huby2012,
   author = {E. Huby and G. Perrin and F. Marchis and S. Lacour and T. Kotani and G. Duchêne and E. Choquet and E. L. Gates and J. M. Woillez and O. Lai and P. Fédou and C. Collin and F. Chapron and V. Arslanyan and K. J. Burns},
   doi = {10.1051/0004-6361/201118517},
   issn = {00046361},
   journal = {Astronomy and Astrophysics},
   title = {\text{FIRST}, a fibered aperture masking instrument: i. First on-sky test results},
   volume = {541},
    pages = {A107},
   year = {2012},
}

@inproceedings{Arcadi2024,
   author = {Elizabeth Arcadi and Glen Douglass and Jacinda Webb and Guillaume Tremblier and Barnaby R. M. Norris and Peter G. Tuthill and Stephanie A. Rossini-Bryson and Eckhart A. Spalding and Marc-Antonie Martinod and Mona El Morsy and Julien Lozi and Vincent Deo and Kyohoon Ahn and Sebastien Vievard and Olivier Guyon and Michael J. Withford and Simon Gross},
   doi = {10.1117/12.3018600},
   isbn = {9781510675230},
   booktitle = {Advances in Optical and Mechanical Technologies for Telescopes and Instrumentation VI},
   month = {8},
   pages = {94},
   publisher = {SPIE},
   title = {Design, fabrication and characterisation of a 3-baseline, achromatic integrated optics beam combiner for nulling interferometry with simultaneous fringe tracking using tricouplers},
volume = {13100},
   year = {2024}
}

@article{Martinod:21,
   author = {Marc-Antoine Martinod and Peter Tuthill and Simon Gross and Barnaby Norris and David Sweeney and Michael J. Withford},
   doi = {10.1364/AO.423541},
   issn = {1559-128X},
   issue = {19},
   journal = {Applied Optics},
   month = {7},
   pages = {D100},
   title = {Achromatic photonic tricouplers for application in nulling interferometry},
   volume = {60},
   url = {https://opg.optica.org/abstract.cfm?URI=ao-60-19-D100},
   year = {2021}
}

@article{Echeverri2024,
   author = {Daniel Echeverri and Jerry W. Xuan and John D. Monnier and Jacques-Robert Delorme and Jason J. Wang and Nemanja Jovanovic and Katelyn Horstman and Garreth Ruane and Bertrand Mennesson and Eugene Serabyn and Dimitri Mawet and J. Kent Wallace and Sofia Hillman and Ashley Baker and Randall Bartos and Benjamin Calvin and Sylvain Cetre and Greg Doppmann and Luke Finnerty and Michael P. Fitzgerald and Chih-Chun Hsu and Joshua Liberman and Ronald López and Maxwell Millar-Blanchaer and Evan Morris and Jacklyn Pezzato and Jean-Baptiste Ruffio and Ben Sappey and Tobias Schofield and Andrew J. Skemer and Ji Wang and Yinzi Xin and Narsireddy Anugu and Sorabh Chhabra and Noura Ibrahim and Stefan Kraus and Gail H. Schaefer and Cyprien Lanthermann},
   doi = {10.3847/2041-8213/ad3619},
   issn = {2041-8205},
   issue = {2},
   journal = {The Astrophysical Journal Letters},
   month = {4},
   pages = {L15},
   title = {Vortex Fiber Nulling for Exoplanet Observations: First Direct Detection of M Dwarf Companions around \text{HIP} 21543, \text{HIP} 94666, and \text{HIP} 50319},
   volume = {965},
   year = {2024}
}

@article{Martinod2021,
   author = {Marc Antoine Martinod and Barnaby Norris and Peter Tuthill and Tiphaine Lagadec and Nemanja Jovanovic and Nick Cvetojevic and Simon Gross and Alexander Arriola and Thomas Gretzinger and Michael J. Withford and Olivier Guyon and Julien Lozi and Sébastien Vievard and Vincent Deo and Jon S. Lawrence and Sergio Leon-Saval},
   doi = {10.1038/s41467-021-22769-x},
   issn = {20411723},
   issue = {1},
   journal = {Nature Communications},
   pmid = {33927206},
   publisher = {Nature Research},
   title = {Scalable photonic-based nulling interferometry with the dispersed multi-baseline GLINT instrument},
pages = {2465},
   volume = {12},
   year = {2021},
}

@article{nasa_exoplanet_archive2025,
  title={The NASA exoplanet archive and exoplanet follow-up observing program: Data, tools, and usage},
  author={Christiansen, Jessie L and McElroy, Douglas L and Harbut, Marcy and Ciardi, David R and Crane, Megan and Good, John and Hardegree-Ullman, Kevin K and Kesseli, Aurora Y and Lund, Michael B and Lynn, Meca and others},
  journal={The Planetary Science Journal},
  volume={6},
  number={8},
  pages={186},
  year={2025},
  publisher={The American Astronomical Society}
}

@article{martinod2023,
  title={High-angular resolution and high contrast observations from Y to L band at the Very Large Telescope Interferometer with the Asgard Instrumental suite},
  author={Martinod, Marc-Antoine and Defr{\`e}re, Denis and Ireland, Michael and Kraus, Stefan and Martinache, Frantz and Tuthill, Peter and Bigioli, Azzurra and Bouzerand, Emilie and Bryant, Julia and Chhabra, Sorabh and others},
  journal={Journal of Astronomical Telescopes, Instruments, and Systems},
  volume={9},
  number={2},
  pages={025007--025007},
  year={2023},
  publisher={Society of Photo-Optical Instrumentation Engineers}
}

@inproceedings{defrere2022,
  title={L-band nulling interferometry at the VLTI with Asgard/Hi-5: status and plans},
  author={Defr{\`e}re, Denis and Bigioli, Azzurra and Dandumont, Colin and Garreau, Germain and Laugier, Romain and Martinod, Marc-Antoine and Absil, Olivier and Berger, Jean-Philippe and Bouzerand, Emilie and Courtney-Barrer, Benjamin and others},
  booktitle={Optical and Infrared Interferometry and Imaging VIII},
  volume={12183},
  pages={184--199},
  year={2022},
  organization={SPIE}
}

@article{Kruse2015,
   author = {Kevin L. Kruse and Christopher T. Middlebrook},
   doi = {10.1080/09500340.2014.983197},
   issn = {0950-0340},
   issue = {sup2},
   journal = {Journal of Modern Optics},
   month = {12},
   pages = {S1-S10},
   title = {Fan-out routing and optical splitting techniques for compact optical interconnects using single-mode polymer waveguides},
   volume = {62},
   year = {2015}
}

@article{Serabyn1999,
author = {E. Serabyn and J. K. Wallace and G. J. Hardy and E. G. H. Schmidtlin and H. T. Nguyen},
journal = {Appl. Opt.},
number = {34},
pages = {7128--7132},
publisher = {Optica Publishing Group},
title = {Deep nulling of visible laser light},
volume = {38},
month = {Dec},
year = {1999},
url = {https://opg.optica.org/ao/abstract.cfm?URI=ao-38-34-7128},
doi = {10.1364/AO.38.007128}
}

@article{Chen2023,
   author = {Hongliang Chen and Guangchen Su and Xin Fu and Lin Yang},
   doi = {10.3390/photonics10121311},
   issn = {2304-6732},
   issue = {12},
   journal = {Photonics},
   month = {11},
   pages = {1311},
   title = {Ultra-Broadband and Compact 2 × 2 3-dB Silicon Adiabatic Coupler Based on Supermode-Injected Adjoint Shape Optimization},
   volume = {10},
   year = {2023}
}

@article{Cheben2018,
   author = {Pavel Cheben and Robert Halir and Jens H. Schmid and Harry A. Atwater and David R. Smith},
   doi = {10.1038/s41586-018-0421-7},
   issn = {0028-0836},
   issue = {7720},
   journal = {Nature},
   month = {8},
   pages = {565-572},
   title = {Subwavelength integrated photonics},
   volume = {560},
   year = {2018}
}

@article{Piggott2015,
   author = {Alexander Y. Piggott and Jesse Lu and Konstantinos G. Lagoudakis and Jan Petykiewicz and Thomas M. Babinec and Jelena Vučković},
   doi = {10.1038/nphoton.2015.69},
   issn = {1749-4885},
   issue = {6},
   journal = {Nature Photonics},
   month = {6},
   pages = {374-377},
   title = {Inverse design and demonstration of a compact and broadband on-chip wavelength demultiplexer},
   volume = {9},
   year = {2015}
}

@ARTICLE{Kojima2021,
  author={Kojima, Keisuke and Tahersima, Mohammad H. and Koike-Akino, Toshiaki and Jha, Devesh K. and Tang, Yingheng and Wang, Ye and Parsons, Kieran},
  journal={Journal of Lightwave Technology}, 
  title={Deep Neural Networks for Inverse Design of Nanophotonic Devices}, 
  year={2021},
  volume={39},
  number={4},
  pages={1010-1019},
  doi={10.1109/JLT.2021.3050083}
}

@article{Hanot2011,
   author = {C. Hanot and B. Mennesson and S. Martin and K. Liewer and F. Loya and D. Mawet and P. Riaud and O. Absil and E. Serabyn},
   doi = {10.1088/0004-637X/729/2/110},
   issn = {0004-637X},
   issue = {2},
   journal = {The Astrophysical Journal},
   month = {3},
   pages = {110},
   title = {IMPROVING INTERFEROMETRIC NULL DEPTH MEASUREMENTS USING STATISTICAL DISTRIBUTIONS: THEORY AND FIRST RESULTS WITH THE PALOMAR FIBER NULLER},
   volume = {729},
   year = {2011}
}

@article{Snellen2015,
   author = {I. Snellen and R. de Kok and J. L. Birkby and B. Brandl and M. Brogi and C. Keller and M. Kenworthy and H. Schwarz and R. Stuik},
   doi = {10.1051/0004-6361/201425018},
   issn = {0004-6361},
   journal = {Astronomy \& Astrophysics},
   month = {4},
   pages = {A59},
   title = {Combining high-dispersion spectroscopy with high contrast imaging: Probing rocky planets around our nearest neighbors},
   volume = {576},
   year = {2015}
}

@article{Douglass2025,
   author = {Glen Douglass and Elizabeth Arcadi and Stephanie Rossini-Bryson and Eckhart Spalding and Marc-Antoine Martinod and Peter Tuthill and Michael J. Withford and Barnaby Norris and Olivier Guyon and Simon Gross},
   doi = {10.1109/JLT.2025.3534653},
   issn = {0733-8724},
   issue = {9},
   journal = {Journal of Lightwave Technology},
   month = {5},
   pages = {4416-4421},
   publisher = {Institute of Electrical and Electronics Engineers Inc.},
   title = {Passive Achromatic Phase Shifter Fabricated Using Ultrafast Laser Inscription},
   volume = {43},
   url = {https://ieeexplore.ieee.org/document/10856357/},
   year = {2025}
}

@article{Ruane2018,
   author = {Garreth Ruane and Ji Wang and Dimitri Mawet and Nemanja Jovanovic and Jacques-Robert Delorme and Bertrand Mennesson and J. Kent Wallace},
   doi = {10.3847/1538-4357/aae262},
   issn = {0004-637X},
   issue = {2},
   journal = {The Astrophysical Journal},
   month = {11},
   pages = {143},
   title = {Efficient Spectroscopy of Exoplanets at Small Angular Separations with Vortex Fiber Nulling},
   volume = {867},
   year = {2018}
}

@ARTICLE{Donnelly1986,
  author={Donnelly, J.},
  journal={IEEE Journal of Quantum Electronics}, 
  title={Limitations on power-transfer efficiency in three-guide optical couplers}, 
  year={1986},
  volume={22},
  number={5},
  pages={610-616},
  doi={10.1109/JQE.1986.1073010}}

@article{Dressing2024Habitable,
	author = {Dressing, Courtney and Ansdell, Megan and Crooke, Julie and Feinberg, Lee and Mennesson, Bertrand and O\textquoteright{}Meara, John and Pepper, Joshua and Roberge, Aki and Ziemer, John and WGs, Habitable Worlds Observatory START, TAG, &},
	journal = {{Bulletin of the AAS}},
	number = {7},
	year = {2024},
	month = {jun 19},
	publisher = {},
	title = {The {Habitable} {Worlds} {Observatory}: Status, {Plans}, and {Opportunities}},
	volume = {56},
}

@article{Sanny2026,
   author = {Ahmed Sanny and Lucas Labadie and Simon Gross and Kévin Barjot and Romain Laugier and Germain Garreau and Marc-Antoine Martinod and Denis Defrère and Michael J. Withford},
   doi = {10.1051/0004-6361/202555865},
   issn = {0004-6361},
   journal = {Astronomy \& Astrophysics},
   month = {1},
   pages = {A37},
   title = {Asgard/NOTT: L-band nulling interferometry at the VLTI},
   volume = {705},
   url = {https://www.aanda.org/10.1051/0004-6361/202555865},
   year = {2026}
}

@article{Gatkine2024,
author = {Pradip Gatkine and Greg Sercel and Nemanja Jovanovic and Ronald Broeke and Katarzyna {\L}awniczuk and Marco Passoni and Ashok Balakrishnan and Serge Bidnyk and Jielong Yin and Jeffrey Jewell and J. Kent Wallace and Dimitri Mawet},
journal = {Opt. Express},
number = {10},
pages = {17689--17703},
publisher = {Optica Publishing Group},
title = {Efficient ultra-broadband low-resolution astrophotonic spectrographs},
volume = {32},
month = {May},
year = {2024},
url = {https://opg.optica.org/oe/abstract.cfm?URI=oe-32-10-17689},
doi = {10.1364/OE.512305},
}

@inproceedings{Bernstein2014,
author = {Rebecca A. Bernstein and Patrick J. McCarthy and Keith Raybould and Bruce C. Bigelow and Atonin H. Bouchez and Jos{\'e} M. Filgueira and George Jacoby and Matt Johns and David Sawyer and Stephen Shectman and Michael Sheehan},
title = {{Overview and status of the Giant Magellan Telescope project}},
volume = {9145},
booktitle = {Ground-based and Airborne Telescopes V},
publisher = {SPIE},
pages = {91451C},
year = {2014},
doi = {10.1117/12.2055282},
URL = {https://doi.org/10.1117/12.2055282}
}

@article{Bryant2017,
   author = {J. J. Bryant and R. R. Thomson and M. J. Withford},
   doi = {10.1364/OE.25.019966},
   issn = {1094-4087},
   issue = {17},
   journal = {Optics Express},
   month = {8},
   pages = {19966},
   title = {Focus issue introduction: recent advances in astrophotonics},
   volume = {25},
   year = {2017}
}

@article{Dinkelaker2021,
   author = {Aline N. Dinkelaker and Aashia Rahman and Joss Bland-Hawthorn and Faustine Cantalloube and Simon Ellis and Philippe Feautrier and Michael Ireland and Lucas Labadie and Robert R. Thomson},
   doi = {10.1364/JOSAB.434565},
   issn = {0740-3224},
   issue = {7},
   journal = {Journal of the Optical Society of America B},
   month = {7},
   pages = {AP1},
   title = {Astrophotonics: introduction to the feature issue},
   volume = {38},
   year = {2021}
}

@inproceedings{KenchingtonGoldsmith2024,
   author = {Harry-Dean Kenchington Goldsmith and Elsa Huby and Manon Lallement and Kévin Barjot and Sébastien Vievard and Vincent Deo and Olivier Guyon and Guillermo Martin and Sylvestre Lacour},
   doi = {10.1117/12.3018270},
   editor = {Stephanie Sallum and Joel Sanchez-Bermudez and Jens Kammerer},
   isbn = {9781510675131},
   booktitle = {Optical and Infrared Interferometry and Imaging IX},
   month = {8},
   pages = {71},
   publisher = {SPIE},
   title = {A new photonic integrated circuit for the FIRST instrument: towards high throughput with a compact photonic chip},
   url = {https://www.spiedigitallibrary.org/conference-proceedings-of-spie/13095/3018270/A-new-photonic-integrated-circuit-for-the-FIRST-instrument/10.1117/12.3018270.full},
   year = {2024}
}

@article{Klinner-Teo2022,
author = {Teresa Klinner-Teo and Marc-Antoine Martinod and Peter Tuthill and Simon Gross and Barnaby Norris and Sergio Leon-Saval},
title = {{Achromatic design of a photonic tricoupler and phase shifter for broadband nulling interferometry}},
volume = {8},
journal = {Journal of Astronomical Telescopes, Instruments, and Systems},
number = {4},
publisher = {SPIE},
pages = {045001},
year = {2022},
doi = {10.1117/1.JATIS.8.4.045001},
URL = {https://doi.org/10.1117/1.JATIS.8.4.045001}
}

@misc{RSOFT,
  author = {{Keysight}},
  title = {{Photonic Design Software}},
  howpublished = {\url{https://www.keysight.com/us/en/products/software/optical-solutions-software/photonic-design-solutions.html}},
  note = {{accessed March 25, 2026}},
}

\end{document}